\documentclass[%
 reprint,
 superscriptaddress,
 amsmath,amssymb,
 aps,
 prb,
 citeautoscript,    %for proper placing of ref. numbers
]{revtex4-2}

\usepackage{graphicx}% Include figure files
\usepackage{dcolumn}% Align table columns on decimal point
\usepackage{bm}% bold math
\usepackage{braket}
\usepackage[bookmarks,colorlinks,
                      linkcolor=blue,
                      filecolor=blue,
                      citecolor=blue,
                       urlcolor=blue]{hyperref}% add hypertext capabilities
\renewcommand{\vec}[1]{\bm{#1}}

\newcommand{\reals}{\mathbb{R}}

\newcommand{\idmatrix}{\mathbb{I}}
\newcommand{\dV}[1]{d^3\vec{#1}}
\newcommand{\dg}[1]{#1^{\dagger}}

\newcommand{\fourierP}[2]{e^{i\vec{#1}\cdot{\vec{#2}}}}

\DeclareMathOperator{\diag}{diag}

\let\epsilon=\varepsilon
\let\phi=\varphi
\let\theta=\vartheta

\usepackage{xcolor}

\begin{document}

%\preprint{APS/123-QED}

\title{Band structures and $\mathbb{Z}_2$ invariants of two-dimensional transition metal dichalcogenide monolayers from fully-relativistic Dirac--Kohn--Sham theory using Gaussian-type orbitals}

\author{Marius Kadek}
\email{marius.kadek@uit.no}
\affiliation{Department of Physics, Northeastern University, Boston, Massachusetts 02115, USA}
\affiliation{Hylleraas Centre for Quantum Molecular Sciences, Department of Chemistry, UiT The Arctic University of Norway, N-9037 Troms\o, Norway}

\author{Baokai Wang}
\affiliation{Department of Physics, Northeastern University, Boston, Massachusetts 02115, USA}

\author{Marc Joosten}
\affiliation{Hylleraas Centre for Quantum Molecular Sciences, Department of Chemistry, UiT The Arctic University of Norway, N-9037 Troms\o, Norway}

\author{Wei-Chi Chiu}
\affiliation{Department of Physics, Northeastern University, Boston, Massachusetts 02115, USA}

\author{Francois Mairesse}   %Francois.mairesse@student.unamur.be
\affiliation{Laboratory of Theoretical Chemistry, University of Namur, B-5000 Namur, Belgium}

\author{Michal Repisky}
\affiliation{Hylleraas Centre for Quantum Molecular Sciences, Department of Chemistry, UiT The Arctic University of Norway, N-9037 Troms\o, Norway}
\affiliation{Department of Physical and Theoretical Chemistry, Faculty of Natural Sciences, Comenius University, Bratislava, Slovakia}

\author{Kenneth Ruud}
\affiliation{Hylleraas Centre for Quantum Molecular Sciences, Department of Chemistry, UiT The Arctic University of Norway, N-9037 Troms\o, Norway}
\affiliation{Norwegian Defence Research Establishment, P.O. Box 25, 2027 Kjeller, Norway}

\author{Arun Bansil}
\affiliation{Department of Physics, Northeastern University, Boston, Massachusetts 02115, USA}


\date{\today}% It is always \today, today,

\begin{abstract}
Two-dimensional (2D) materials exhibit a wide range of remarkable phenomena, many of which owe their existence to the relativistic spin--orbit coupling (SOC) effects. To understand and predict properties of materials containing heavy elements, such as the transition-metal dichalcogenides (TMDs), relativistic effects must be taken into account in first-principles calculations. We present an all-electron method based on the four-component Dirac Hamiltonian and Gaussian-type orbitals (GTOs) that overcomes complications associated with linear dependencies and ill-conditioned matrices that arise when diffuse functions are included in the basis. Until now, there has been no systematic study of the convergence of GTO basis sets for periodic solids either at the nonrelativistic or the relativistic level. Here we provide such a study of relativistic band structures of the 2D TMDs in the hexagonal (2H), tetragonal (1T), and distorted tetragonal (1T') structures, along with a discussion of their SOC-driven properties (Rashba splitting and $\mathbb{Z}_2$ topological invariants). We demonstrate the viability of our approach even when large basis sets with multiple basis functions involving various valence orbitals (denoted triple- and quadruple-$\zeta$) are used in the relativistic regime. Our method does not require the use of pseudopotentials and provides access to all electronic states within the same framework. Our study paves the way for direct studies of material properties, such as the parameters in spin Hamiltonians, that depend heavily on the electron density near atomic nuclei where relativistic and SOC effects are the strongest.
\end{abstract}

\maketitle

%\tableofcontents

\section{\label{sec:Introduction}Introduction}

Two-dimensional (2D) materials~\cite{butler2013progress,han2016perspectives} are solids in which atoms or compounds are bound together by strong bonds (\emph{e.g.} covalent or ionic) along two spatial dimensions, confining electron transport to a plane. In the out-of-plane dimension, weaker (van der Waals) forces enable the synthesis of materials with thicknesses of only a few atomic layers~\cite{frindt1966single,geim2007rise,sarma2011electronic,miro2014atlas}. 2D materials have recently become very attractive as they exhibit remarkable transport~\cite{tombros2007electronic,li2013intrinsic}, topological~\cite{kane2005quantum,liu2011quantum,qian2014quantum}, thermoelectric~\cite{hor2009p}, and optoelectronic~\cite{wang2012electronics} properties that can be exploited to develop novel devices for quantum computing~\cite{nilsson2008splitting,fu2009josephson}, field-effect transistors~\cite{radisavljevic2011single,britnell2012field}, low-power logic devices~\cite{solomon2015pathway}, and strain-controllable light-emitting devices~\cite{li2015optoelectronic}. Advantages of 2D materials over conventional solids stem from their thin surfaces which enable considerable manipulation of their properties that can be controlled using defects, adatoms, and electric-field gating~\cite{brar2011gate,natterer2012ring,tsai2013gated}. In addition, atomically thin materials can be manually assembled to form multilayered heterostructures~\cite{ponomarenko2011tunable,geim2013van,britnell2012field,qian2014quantum} to combine functionalities of individual layers~\cite{haigh2012cross}.

Monolayers of transition-metal dichalcogenides (TMDs)~\cite{manzeli20172d} are 2D materials of type $MX_2$, where $M$ is a transition-metal atom (Mo, W, \ldots), and $X$ is a chalcogen atom (S, Se, Te), and constitute basic building blocks for many heterostructures. TMDs host many exotic physical phenomena, such as the quantum spin Hall (QSH)~\cite{qian2014quantum} and nonlinear anomalous Hall~\cite{kang2019nonlinear} effects, higher-order topology~\cite{wang2019higher}, giant Rashba spin-splittings of valence bands~\cite{zhu2011giant}, as well as various correlated phases, \emph{e.g.} charge density waves~\cite{wilson1974charge,flicker2015charge}, superconductivity and ferromagnetism~\cite{zhu2016signature}. TMDs typically contain elements from the lower part of the periodic table, where the relativistic theory of electrons is unavoidable. Many interesting properties of the TMDs owe their existence to the relativistic link between the spin and orbital degrees of freedom, \emph{i.e.} the spin--orbit coupling (SOC) effects. The ability to control and tune SOC in TMDs opens new possibilities for spintronics and valleytronics devices based on non-magnetic materials, where spin is manipulated by electrical means only~\cite{xiao2012coupled,mak2014valley,pulkin2016spin,xu2014spin}. Furthermore, SOC generates opposite effective Zeeman fields at the $K$ and $K$' valleys of the TMDs, which enables formation of Cooper pairs and opens the possibility of observing topological superconductivity and Majorana fermions~\cite{zhou2016ising,he2018magnetic}. Recently, antisite defects in TMDs were proposed to be suitable for hosting solid-state spin qubits, where SOC enables transitions between different spin configurations required for qubit operations~\cite{tsai2022antisite}.

First-principles computational approaches enable systematic, parameter-free, material-specific predictions in novel materials. However, due to the presence of heavy elements in the TMDs, it is important that relativistic and SOC effects are accounted for in the theoretical framework. In this connection, the four-component (4c) Dirac Hamiltonian~\cite{dirac1928quantum,saue2002four,belpassi2011recent}
\begin{equation}
	\hat{H} = \begin{pmatrix}
		V(\vec{r}) & c\vec{\sigma}\cdot\vec{\pi} \\
		c\vec{\sigma}\cdot\vec{\pi} & V(\vec{r}) - 2c^2
	\end{pmatrix},
\label{eq:DiracHamiltonian}
\end{equation}
is the commonly accepted gold standard, which is exact in the limit of noninteracting relativistic electrons. Here, $\vec{\sigma}$ are the Pauli matrices, $\vec{\pi}\equiv-i\vec{\nabla} + \vec{A}$ is the canonical momentum operator of the electron, $c$ is the speed of light, and atomic units ($e = \hbar = m_e = 1$) are used. The scalar and vector potentials $V$ and $\vec{A}$, respectively, describe the interaction of the electron with external electromagnetic fields, where $V$ typically contains interactions with the nuclear charges. The lower components of the Dirac bispinor wave functions are associated with the negative-energy states, and various strategies for their elimination lead to a number of approximate two-component (2c) schemes~\cite{saue2011relativistic,liu2014advances}.

The multiple wave-function components needed in relativistic simulations significantly increase the computational cost as larger matrix and vector dimensions are involved. Kohn--Sham (KS) density functional theory (DFT)~\cite{hohenberg1964inhomogeneous,kohn1965self} brings relativistic calculations to an affordable level and remains the method of choice for large solid-state systems containing heavy elements. However, we note the work of Yeh \emph{et al.}~\cite{yeh2022relativistic} on the self-consistent $GW$ method at the relativistic 2c level of theory. In KS theory, the potential in Eq.~\eqref{eq:DiracHamiltonian} also contains terms that depend on the electron density and its gradient: the mean-field Coulomb potential generated by other electrons, and the exchange--correlation potential~\cite{rajagopal1973inhomogeneous,saue2002four,komorovsky2008fully,chapterEngel2002,dyall2007introduction}. The relativistic corrections to the instantaneous electron--electron Coulomb interactions are here neglected. If SOC is treated self-consistently, expensive KS Hamiltonian matrix constructions must be carried out in the multicomponent regime. As a consequence, most internal parts of quantum mechanical codes must be reconsidered and adapted. A transparent theoretical formalism that isolates the SOC terms, reduces computational cost, and simplifies the implementation of fully relativistic theories can be achieved using quaternion algebra~\cite{saue1997principles,konecny2018resolution,kadek2019all,repisky2020respect}.

The most commonly applied strategies incorporate SOC in relativistic pseudopotentials and the related projector-augmented wave methods~\cite{kleinman1980relativistic,bachelet1982relativistic,dal2005spin,nichols2009gaussian,dal2010projector}, where the oscillating wave function with complicated nodal structure in the core region close to the nuclei is replaced by a smooth pseudo-wavefunction. However, all-electron approaches are necessary to describe Rashba-like spin-splitting induced by distortions of the wave function close to the nucleus~\cite{bihlmayer2006rashba}, where nuclear spins interact with the electrons~\cite{ghosh2021spin}, or in situations where high accuracy is desired~\cite{gmitra2009band,sichau2019resonance}. Linearized augmented plane-wave (LAPW) ~\cite{andersen1975linear,macdonald1980linearised,wimmer1981full} and full-potential linear muffin-tin orbital (LMTO)~\cite{christensen1978spin,ebert1988spin,rehn2020dirac} methods enable all-electron calculations by constructing muffin-tin spheres around all atoms and expressing the all-electron wave functions using orbitals inside these spheres and plane waves in the interstitial region. In such cases, treatment of SOC must be handled separately in the two distinct regions and is sometimes neglected outside of the muffin-tin spheres. An alternative all-electron approach to LAPW and LMTO can be formulated using a finite-element basis set~\cite{motamarri2020dft} or by expanding the Bloch wave functions $\psi_n(\vec{k};\vec{r})$ using linear combinations of atom-centered real-space basis functions $\chi_{\mu}(\vec{r})$. Techniques that employ various choices for $\chi_{\mu}$, such as the numerical~\cite{suzuki1999fully,koepernik1999full,zhao2021quasi}, Slater-type~\cite{philipsen1997relativistic,zhao2016exact}, or Gaussian-type basis set~\cite{boettger2007first,kadek2019all}, have been developed for periodic systems at the 4c~\cite{suzuki1999fully,koepernik1999full,kadek2019all} as well as the approximate relativistic~\cite{boettger2007first,philipsen1997relativistic,zhao2016exact,zhao2021quasi} levels of the theory.

In the 4c theory, the basis functions $\chi_{\mu}(\vec{r})$ need to represent Dirac bispinors, and the basis expansion of the 4c Bloch wave functions $\psi_n(\vec{k};\vec{r})$ takes the form:
\begin{equation}
	\psi_n(\vec{k};\vec{r}) = \frac{1}{\sqrt{|\mathcal{K}|}} \sum_{\vec{R},\mu} \fourierP{k}{R} \chi_{\mu}(\vec{r}-\vec{R}) c^{\mu}_n(\vec{k}),
\label{eq:BlochFunctions}
\end{equation}
where $n$ is the band index, $\vec{k}$ is the reciprocal-space vector (quasi-momentum), $\vec{R}$ denotes the Bravais lattice vector of the respective unit cell in the translationally invariant system ($\vec{R}=\vec{0}$ is the reference unit cell), index $\mu$ runs over the scalar basis functions in the unit cell, and the normalization constant is the inverse square root of the volume of the primitive reciprocal unit cell~$|\mathcal{K}|$. The expansion coefficients $c^{\mu}_n(\vec{k})$ and the ground-state electronic structure are obtained by solving the matrix form of the reciprocal-space Dirac--Kohn--Sham (DKS) generalized eigenvalue equation
\begin{equation}
	H(\vec{k})c(\vec{k}) = S(\vec{k})c(\vec{k})\epsilon(\vec{k}),
\label{eq:DKSinMatrixForm}
\end{equation}
where $\epsilon(\vec{k})$ is the diagonal matrix of eigenvalues (band energies), and $H(\vec{k})$ and $S(\vec{k})$ are the reciprocal-space DKS Hamiltonian and overlap matrices, respectively, with the elements:
\begin{subequations}
\label{eq:kSpaceGroupFandS}
\begin{align}
	H_{\mu\mu'}(\vec{k}) &= \sum_{\vec{R}} \fourierP{k}{R} \int_{\reals^3} \dg{\chi}_{\mu}(\vec{r}) \hat{H}\chi_{\mu'}(\vec{r}-\vec{R}) \dV{r}, \label{eq:kSpaceFock}\\
	S_{\mu\mu'}(\vec{k}) &= \sum_{\vec{R}} \fourierP{k}{R} \int_{\reals^3} \dg{\chi}_{\mu}(\vec{r}) \chi_{\mu'}(\vec{r}-\vec{R}) \dV{r}. \label{eq:kSpaceOverlap}
\end{align}
\end{subequations}
Due to the dependence of $H(\vec{k})$ on the electron density and its gradient that are determined from $c(\vec{k})$, Eq.~\eqref{eq:DKSinMatrixForm} must be solved self-consistently, though the SOC terms are often neglected in the self-consistency procedure and only included \emph{a posteriori} as a perturbation to save computational cost~\cite{huhn2017one}.

Our goal here is to provide an all-electron approach for modeling heavy-element-containing 2D materials based on the self-consistent treatment of SOC and Gaussian-type orbitals (GTOs) commonly used in quantum chemistry calculations of properties, electronic structures, and electron correlations in molecules. The use of GTOs for solids offers several advantages. For instance, a unified representation of wave functions in a variety of systems (molecules, polymers, 2D materials, crystals) allows for building on the existing quantum chemistry approaches, such as the accurate treatment of correlation~\cite{pisani2012cryscor,booth2016plane,mcclain2017gaussian,rebolini2018divide}, reduced cost of evaluating the exact exchange~\cite{crowley2016resolution}, and efficient algorithms that scale linearly with system size~\cite{kudin2000linear,lazarski2015density}. In 2D materials that are only a few atomic layers thick, Bloch functions are straightforwardly constructed to satisfy the Bloch theorem only across the two periodic dimensions by explicitly restricting $\vec{R}$ and $\vec{k}$ in Eq.~\eqref{eq:BlochFunctions} to 2D lattice vectors. As a consequence, the system is not artificially replicated in the nonperiodic dimension. For comparison with calculations employing plane waves, this situation would correspond to the limit of infinitely large vacuum layers (or zero hopping) between the replicated images. Finally, explicit self-consistent treatment of SOC for all electronic states on an equal footing without adopting the muffin-tin or pseudopotential approximations provides a pathway for modeling x-ray absorption spectroscopy~\cite{kadek2015x} and magnetic response properties associated with nuclear spins~\cite{komorovsky2006resolution,repisky2009restricted}.

Modelling solid-state systems with GTOs in the nonrelativistic framework was pioneered by Pisani and Dovesi~\cite{pisani1980exact,dovesi1983treatment,dovesi2020crystal}, later joined by several other groups~\cite{kudin2000linear,lazarski2015density,sun2020recent}. However, standard GTO basis sets are constructed by optimizing calculations on atoms and thus contain basis functions with small exponents to describe the asymptotic behavior of the atomic wave functions. These diffuse functions severely hamper the application of GTOs to solids and cause numerical instabilities~\cite{suhai1982error,kudin2000linear,peintinger2013consistent,zhu2021all}. The numerical issues can be circumvented by removing the most diffuse basis functions at the start of the calculation~\cite{peralta2005scalar,usvyat2011approaching,perry2001antiferromagnetic} or by constructing system-specific basis sets using reoptimized Gaussian exponents and contraction coefficients~\cite{peintinger2013consistent,daga2020gaussian}. The former strategy risks producing low quality results~\cite{kadek2019all,jensen2013analysis} whereas the latter strategy sacrifices the transferability of the so-constructed basis sets and requires availability of advanced code features (basis set optimization). Pre-optimized all-electron basis sets are also not available for heavy elements~\cite{peintinger2013consistent}. The importance of numerical stability is even more pronounced in relativistic theories where the additional wave-function components increase the variational freedom, and the restricted kinetic balance (RKB) condition~\cite{stanton1984kinetic} must be satisfied to prevent the collapse of the spectral gap between the positive- and negative-energy states.

In this paper, we demonstrate that accurate and converged results can be obtained for 2D TMD monolayers in various structural phases using commonly available all-electron valence triple-$\zeta$ basis sets in the fully relativistic framework without the need for modifying or reoptimizing the basis functions. Our approach builds on the quaternion algebra-based theory~\cite{kadek2019all} implemented here with careful numerical considerations to ensure robustness of the implementation and the resolution-of-the-identity (RI) approximation for the Coulomb four-center integrals ~\cite{baerends1973self,jung2005auxiliary,varga2008long} that reduces the computational cost of otherwise time-consuming calculations by more than three orders of magnitude. We show how the canonical orthogonalization can be used to construct numerically well-behaved orthonormal bases in momentum space in the 4c setting and derive a criterion for assessing the quality of the RKB condition in such bases. We discuss a parallel algorithm for the electronic structure solver exhibiting minimal input/output (IO) disk communication and memory overhead that is applicable to unit cells with several thousand basis functions and thousands of $\vec{k}$ points. The method presented here is implemented in the \textsc{ReSpect} code~\cite{repisky2020respect} and used to calculate the band structures of the hexagonal (2H), tetragonal (1T), and distorted tetragonal (1T') phases of selected TMDs. For the 1T' phase, we evaluate the $\mathbb{Z}_2$ invariant within our real-space GTO-based scheme and confirm the findings of Qian \emph{et al.}~\cite{qian2014quantum} that the 1T' phase of $MX_2$ is topologically nontrivial.

The data presented in this study as well as the scripts used for pre- and post-processing of the input and output files are available in the \href{https://doi.org/10.5281/zenodo.7394905}{ZENODO public repository}~\cite{dataset}. The \textsc{ReSpect} code used in this study is available upon a reasonable request free of charge, see Ref.~\cite{respect}.

\section{\label{sec:Methods}Methods}

\subsection{\label{sec:Construction}Construction of orthonormal momentum-space basis}

Throughout this section, we assume Einstein's implicit summation over doubly repeating indices.

In order to avoid variational collapse associated with an incomplete basis representation of the lower components of the wave function~\cite{schwarz1982basis} and to obtain the correct nonrelativistic limit of the kinetic energy operator in a finite basis, we impose the RKB condition~\cite{stanton1984kinetic} for the basis bispinors $\chi_{\mu}(\vec{r})$ in real space. Hence, we require that~\cite{dyall1994exact,komorovsky2008fully,repisky2015excitation,konecny2018resolution,repisky2020respect,saue2011relativistic}
\begin{equation}
	\chi_{\mu}(\vec{r}) \equiv \begin{pmatrix}
		\idmatrix_2 & 0_2 \\
		0_2 & \frac{1}{2c}\vec{\sigma}\cdot\vec{p}
	\end{pmatrix}g_{\mu}(\vec{r}-\vec{A}_{\mu}),
\label{eq:AObasis4c}
\end{equation}
where $\idmatrix_2$ is the $2\times 2$ identity matrix, $\vec{p}\equiv-i\vec{\nabla}$ is the electron momentum operator, and $g_{\mu}$ are scalar basis functions centered on atomic positions $\vec{A}_{\mu}$. Due to the multicomponent structure of $\hat{H}$ and $\chi_{\mu}$, $H_{\mu\mu'}$ and $S_{\mu\mu'}$ in Eq.~\eqref{eq:kSpaceGroupFandS} are $4\times 4$ complex matrices for each pair $\mu,\mu'$ and for each $\vec{k}$. For the functions $g_{\mu}$, we choose the primitive spherical GTOs
\begin{equation}
	g_{\mu}(\vec{r}) \equiv \mathcal{N} Y_{lm}(\theta,\phi) e^{-\alpha\vec{r}^2},
\label{eq:sphericalGTOs}
\end{equation}
where $\mathcal{N}$ is the normalization constant, $\alpha$ is the Gaussian exponent, and $Y_{lm}(\theta,\phi)$ are the spherical harmonics. The GTO basis is commonly implemented in many quantum chemistry codes~\cite{dovesi2020crystal,balasubramani2020turbomole,apra2020nwchem,saue2020dirac,olsen2020dalton,belpassi2020bertha,sun2020recent}. For computational reasons, integrals in Eqs.~\eqref{eq:kSpaceGroupFandS} are evaluated analytically by solving recurrence relations~\cite{obara1986efficient} formulated in terms of the Cartesian GTOs~\cite{boys1950electronic}. The resulting integrals are then transformed to spherical GTOs.

To construct an orthonormal basis in momentum space, let us first define a nonorthogonal basis from $\chi_{\mu\vec{R}}(\vec{r})\equiv\chi_{\mu}(\vec{r}-\vec{R})$ as
\begin{equation}
	\chi_{\mu}(\vec{k}) = \frac{1}{\sqrt{|\mathcal{K}|}} \sum_{\vec{R}} \fourierP{k}{R} \chi_{\mu\vec{R}}.
\label{eq:BlochBasis}
\end{equation}
A new basis is obtained using the transformation:
\begin{equation}
	\phi_p(\vec{k}) = \chi_{\mu}(\vec{k})B^{\mu}_p(\vec{k}).
\label{eq:basis2ONO}
\end{equation}
From here, for clarity, we drop the dependence of all matrices on $\vec{k}$. The matrix $B$ can be chosen so that $\phi_p$ are orthonormal, \emph{i.e.} $\dg{B}SB = \idmatrix$. We perform the canonical orthogonalization,
\begin{equation}
	B = Us^{-1/2}\Lambda,
\label{eq:canonicalOGO}
\end{equation}
where $U$ is the unitary matrix of eigenvectors and $s$ is the diagonal matrix of eigenvalues, both obtained by diagonalizing the overlap matrix $S$. $\Lambda$ is a rectangular matrix consisting of the square identity matrix and zero rows that correspond to the basis functions that are projected out. In case no projections are required, $\Lambda$ is simply the identity matrix.

In the relativistic 4c theory, $\chi_{\mu}$ take the bispinor form of Eq.~\eqref{eq:AObasis4c}, and the $B^{\mu}_p$ become $4\times 4$ matrices for each $\mu,p$. However, let us now consider a general basis set of bispinors $\chi_{\mu}$ with $\chi^L_{\mu}$ and $\chi^S_{\mu}$ denoting the $2\times 2$ large (upper) and small (lower) components of the basis, respectively. Similarly, let $\phi^L_p\equiv\chi^L_{\mu}X^{\mu}_p$ and $\phi^S_p\equiv\chi^S_{\mu}Y^{\mu}_p$ denote the transformed (\emph{e.g.} orthonormalized) large and small components of the basis, respectively, $X$ and $Y$ forming a block-diagonal matrix $B$ in Eq.~\eqref{eq:basis2ONO}. The DKS Hamiltonian takes the matrix form:
\begin{equation}
	H = \begin{pmatrix}
		V_{LL} & c\Pi_{LS} \\
		c\Pi_{SL} & V_{SS} - 2c^2 S_{SS}
	\end{pmatrix},
\label{eq:Fock4cOperatorBasis}
\end{equation}
where $V_{LL}, V_{SS}, \Pi_{LS}, \Pi_{SL}$ are the matrix representations of their respective operators in the $\chi_{\mu}$ basis, and $S_{SS}$ is the overlap matrix for the small-component basis. Ensuring numerical stability requires projecting out basis functions $\phi^L_p$ and $\phi^S_p$ corresponding to the smallest eigenvalues of the 4c overlap matrix. At the same time, removing $\phi^S_p$ functions degrades the RKB condition that is no longer satisfied exactly in the new basis, which can lead to the emergence of artificial in-gap states. Here, we proceed by showing how we numerically track this basis truncation error. First, we express the RKB condition in the matrix form using the arbitrary basis $\chi_{\mu}$. If we write the 4c eigenvalue problem with the Hamiltonian matrix in Eq.~\eqref{eq:Fock4cOperatorBasis} and eliminate the small-component wave-function coefficients, we obtain
\begin{equation}
	\left[V_{LL} + \frac{1}{2}\Pi_{LS}\mathcal{B}^{-1}_{SS}(\epsilon)\Pi_{SL}\right] c_L = \epsilon S_{LL} c_L,
\label{eq:2cExact}
\end{equation}
where
\begin{equation}
	\mathcal{B}_{SS}(\epsilon) = S_{SS} + \frac{\epsilon S_{SS} - V_{SS}}{2c^2},
\label{eq:2cBoper}
\end{equation}
$c_L$ are the coefficients of the large component of the wave function, and $\epsilon$ is the one-electron energy. Imposing the requirement that the nonrelativistic Hamiltonian is recovered in the limit $c\rightarrow\infty$ gives the following condition:
\begin{equation}
	T_{LL} = \frac{1}{2} \Pi_{LS}S^{-1}_{SS}\Pi_{SL},
\label{eq:RKBgeneral}
\end{equation}
where $T_{LL}$ is the nonrelativistic kinetic energy matrix in the basis of the large component. We note that the choice of bispinors $\chi_{\mu}$, as in Eq.~\eqref{eq:AObasis4c}, satisfies Eq.~\eqref{eq:RKBgeneral} analytically, since $\Pi_{LS} = \Pi_{SL} = \frac{1}{c}T_{LL}$ and $S_{SS} = \frac{1}{2c^2}T_{LL}$. In the transformed basis $\phi_p$, Eq.~\eqref{eq:RKBgeneral} reads:
\begin{equation}
	\dg{X}T_{LL}X = \frac{1}{2} \dg{X}\Pi_{LS}Y\left(\dg{Y}S_{SS}Y\right)^{-1}\dg{Y}\Pi_{SL}X.
\label{eq:RKBtransformed}
\end{equation}
This RKB condition remains valid as long as two conditions are met: The former basis $\chi_{\mu}$ satisfies Eq.~\eqref{eq:RKBgeneral}, and the inverse $Y^{-1}$ exists, \emph{i.e.} the new basis spans the same space as the original one. By requiring that the new basis is orthonormal ($\dg{Y}S_{SS}Y = \idmatrix$), this condition simplifies to:
\begin{equation}
	\dg{X}T_{LL}X = \frac{1}{2} \dg{X}\Pi_{LS}Y\dg{Y}\Pi_{SL}X,
\label{eq:RKBtransformedONO}
\end{equation}
and the existence of $Y^{-1}$ implies that $S^{-1}_{SS} = Y\dg{Y}$. However, orthonormalization procedures that involve removing eigenvectors in the orthonormal basis of $\phi^S_p$ by using a rectangular $\Lambda$ in Eq.~\eqref{eq:canonicalOGO} do not generally preserve the RKB condition. Hence, in the 4c framework, improving the conditioning of matrices by projecting out redundant basis functions can sacrifice the RKB between the $L$ and $S$ components.

To assess the quality of the orthonormal basis for the lower components of the wave function, we calculate how much of the RKB condition is lost by the transformation to the reduced orthonormal basis. To this end, we evaluate
\begin{equation}
    e = \max\left[ \dg{X}\left(\frac{1}{2c^2} T_{LL}Y\dg{Y}T_{LL} - T_{LL}\right)X\right],
\label{eq:RKBtest}
\end{equation}
for each $\vec{k}$. Here, $\max$ indicates the largest matrix element. We found that large values of $e$ ($>10^{-5}$) could be associated with the emergence of artificial states in the band structure. To keep the error in the RKB small, for every $\phi^S_p$ function that is projected out, we also remove the corresponding $\phi^L_p$ function regardless of how large the eigenvalue of the overlap matrix $S_{LL}$ is for this function. This means that the $\phi^L_p$ must be removed even if it is not causing overcompleteness of the large-component subspace. We note that these measures are necessary for moderately sized (triple-$\zeta$ and quadruple-$\zeta$) basis sets, but the numerical issues are not observed in case the smaller double-$\zeta$ basis are used. Finally, the process of orthonormalization and removal of redundant functions heavily depends on $\vec{k}$, so the implementation of the electronic structure solver must have the flexibility to account for matrices with different sizes for each $\vec{k}$.

\subsection{\label{sec:Parallelized}Parallel electronic structure solver}

Construction of the ground-state one-electron wave functions within the self-consistent field (SCF) theory consists of two distinct steps that are iterated repeatedly until self-consistency is reached. The DKS Hamiltonian and overlap matrices in Eqs.~\eqref{eq:kSpaceGroupFandS} are first assembled in real space using the GTO basis. These are then transformed to reciprocal space, and Eq.~\eqref{eq:DKSinMatrixForm} is solved in the orthonormal basis using matrix algebra (multiplications and diagonalizations). Large-scale or high-throughput calculations of band structures of solids require efficient algorithms that offer good (ideally linear) scaling with respect to the number of computer cores used as well as the size of the unit cell. This must be accomplished for both of the SCF solver steps. The demand for efficiency is especially important for all-electron 4c calculations, where the matrix sizes are much larger and even otherwise negligible matrix operations become time-consuming.

Evaluation of the DKS Hamiltonian matrix elements is dominated by the electron--electron Coulomb contributions for DFT simulations with pure exchange--correlation functionals. We adopt the resolution-of-the-identity (RI) approach~\cite{baerends1973self,jung2005auxiliary,varga2008long} (also known as the density fitting procedure) combined with multipole expansions~\cite{sierka2003fast,lazarski2015density} to significantly reduce the overall cost of computing the Coulomb terms. The RI method approximates the electron density by a linear combination of auxiliary basis functions centered on atoms, as opposed to the exact treatment of the four-center Coulomb integrals where the orbital products that constitute the density are expanded individually. Our RI implementation differs from earlier work~\cite{lazarski2015density,sun2017gaussian}. The theoretical challenges associated with the divergent terms in the Coulomb metric matrix of periodic systems together with details of our RI method will be presented elsewhere.

Once the real-space matrix elements are calculated, the remaining steps -- transformation to reciprocal space, orthonormalization procedure, matrix diagonalization -- is performed independently for each $\vec{k}$. We exploit factorization of the reciprocal-space tasks by employing message-passing-interface (MPI) directives which allow high level of parallelization to be achieved due to almost no communication needed among various MPI processes. Each process utilizes multiple OpenMP threads when calling internally parallelized matrix libraries. However, in order to construct the electron density,
\begin{equation}
    \rho(\vec{r}) = \sum_n \int_{\mathcal{K}} f_n(\vec{k})\dg{\psi}_n(\vec{k};\vec{r})\psi_n(\vec{k};\vec{r}) \dV{k},
\label{eq:density}
\end{equation}
where $f_n(\vec{k})$ is the occupation number of the $n$-th band and $\psi_n(\vec{k};\vec{r})$ are the Bloch functions defined in Eq.~\eqref{eq:BlochFunctions}. Band energies $\epsilon_n(\vec{k})$, from which the occupations are determined, must be known for all $\vec{k}$. Hence, the evaluation of the electron density (needed for subsequent SCF iterations) occurs after all the wave-function coefficients $c^{\mu}_n(\vec{k})$ and energies are found. In the all-electron 4c framework, the coefficient matrices are too large to be stored in memory for all $\vec{k}$ and storing the matrices on disk hampers the MPI parallelization due to significantly increased I/O communication. To this end, we partition the electron density into two terms:
\begin{equation}
    \rho(\vec{r}) = \rho_\text{c}(\vec{r}) + \rho_\text{v}(\vec{r}),
\label{eq:densitySplit}
\end{equation}
where, for $\rho_\text{c}$, we restrict the sum $\sum_n$ in Eq.~\eqref{eq:density} to contain only those bands that are assumed to be fully filled ($f_n(\vec{k})=1$) during the entire SCF procedure. $\rho_\text{c}$ contains contributions to the electron density from the majority of bands -- these terms are evaluated on-the-fly after diagonalization independently for each $\vec{k}$ point without the need to store the full coefficient matrices. The remaining terms in the density are incorporated in $\rho_\text{v}$ and calculated in a separate loop over $\vec{k}$ after all $\epsilon_n(\vec{k})$ and $f_n(\vec{k})$ are known, which allows for a small number of partially filled bands [$f_n(\vec{k})$ to be varied across $\vec{k}$]. The coefficients $c^{\mu}_n(\vec{k})$ needed for $\rho_\text{v}$ can be stored in memory as they form narrow rectangular matrices with only one dimension increasing with the system size. In addition, each MPI process only needs to keep the coefficients for the subset of $\vec{k}$ points that the process handles, which is advantageous for multinode calculations within the distributed-memory architectures since the coefficients do not need to be shared or communicated among nodes. The parallel scheme described here enables seamless all-electron fully relativistic 4c calculations of band structures and density-of-states with a large number of $\vec{k}$ points using thousands of cores.

\subsection{\label{sec:Evaluation}Evaluation of $\mathbb{Z}_2$ invariant}

The QSH phase of 2D materials with time-reversal symmetry is characterized by a nontrivial topological order associated with a nonzero $\mathbb{Z}_2$ index~\cite{kane2005z}. Identification of topological materials from numerical simulations is not straightforward~\cite{gresch2017z2pack}. However, for systems with inversion symmetry, the calculation of the $\mathbb{Z}_2$ invariant can be simplified using Fu and Kane's method~\cite{fu2007topological} based on the knowledge of the parity of the Bloch wave functions of the bulk crystal at the time-reversal invariant momenta (TRIM) of the Brillouin zone. In this section, we describe how the $\mathbb{Z}_2$ invariant is calculated in the 4c real-space GTO basis.

First, we discuss the construction of the momentum-space parity operator $P_{\mu\mu'}(\vec{k})$. Let $\mathcal{I}$ denote the space-inversion operator with the inversion center at $\vec{G}$, defined by its action on a scalar function $f$ as $\mathcal{I}f(\vec{r})\equiv f(2\vec{G}-\vec{r})$. From the requirement that the parity-transformed bispinor wave function $\psi(\vec{r},t)$ satisfies the same Dirac equation, it is possible to identify the form of the 4c parity operator $\hat{P}$ as:
\begin{equation}
    \hat{P}\psi(\vec{r},t) = \eta\beta \psi(2\vec{G}-\vec{r},t) \equiv \eta\beta\mathcal{I} \psi(\vec{r},t),
\label{eq:parityOnBispinor}
\end{equation}
where $\beta = \diag(\idmatrix_2,-\idmatrix_2)$ and $\eta$ is an arbitrary phase factor ($|\eta|=1$) that can be introduced by the parity transformation. Neglecting the phase gives $\hat{P} = \beta\mathcal{I}$. The 4c matrix form of the parity operator is obtained by realizing that $\mathcal{I}\vec{p}\dg{\mathcal{I}} = -\vec{p}$ and letting $\hat{P}$ act on the RKB basis in Eq.~\eqref{eq:AObasis4c}, \emph{i.e.}
\begin{equation}
	\hat{P}\chi_{\mu} \equiv \hat{P} \begin{pmatrix}
		\idmatrix_2 & 0_2 \\
		0_2 & \frac{1}{2c}\vec{\sigma}\cdot\vec{p}
	\end{pmatrix}g_{\mu} = \begin{pmatrix}
		\idmatrix_2 & 0_2 \\
		0_2 & \frac{1}{2c}\vec{\sigma}\cdot\vec{p}
	\end{pmatrix}\mathcal{I}g_{\mu}.
\label{eq:parityOpRKB}
\end{equation}
It follows that the parity matrix $P_{\mu\mu'}(\vec{k})$ can be obtained using Eq.~\eqref{eq:kSpaceOverlap} for the overlap matrix with an additional application of the inversion $\mathcal{I}$ on the scalar basis function $g_{\mu'}$. For an orbital from Eq.~\eqref{eq:sphericalGTOs} with angular momentum number $l$ that is centered at $\vec{A}$ (which here denotes the position of an atom $\vec{A}_{\mu}$ in a unit cell $\vec{R}$), this inversion gives:
\begin{equation}
    \mathcal{I}g_{\mu}(\vec{r}-\vec{A}) = (-1)^l g_{\mu}\left(\vec{r}+\vec{A}-2\vec{G}\right).
\label{eq:parityOnScalarBasis}
\end{equation}
Finally, the $\mathbb{Z}_2$ invariant is calculated using the eigenvalues $\xi_{2m}$ of the parity $P_{\mu\mu'}(\vec{k})$ as~\cite{fu2007topological}:
\begin{equation}
    (-1)^{\nu} = \prod_{i=1}^4 \prod_{m=1}^N \xi_{2m}(\Gamma_i),
\label{eq:Z2index}
\end{equation}
where $N$ is the number of occupied Kramers pairs, and $\Gamma_i$ labels the four TRIM points in 2D.

We conclude this section by noting a few considerations required for a numerically stable and reliable calculation of the parity matrix and its eigenvalues when using the method described above. First, the sparse matrix storage scheme outlined in Sec.~\ref{sec:Construction} based on the locality of basis function products $\dg{\chi}_{\mu}(\vec{r}) \chi_{\mu'}(\vec{r}-\vec{R})$ needs to be modified to respect the nonlocal nature of the inversion operator. Specifically, the list of significant elements of the inversion matrix is not the same as for the overlap matrix but it is related via a reflection. Likewise, it must be ensured that the atomic basis set chosen for the inversion-equivalent atoms is the same. Finally, it is not automatically guaranteed that the Bloch functions in Eq.~\eqref{eq:BlochFunctions} that are solutions of the DKS equation are also eigenfunctions of the parity operator for Bloch states with degenerate energies. Parity eigenvalues can be obtained by diagonalizing the sub-blocks of the parity matrix that correspond to these degenerate energy levels. However, in numerical simulations, energy values obtained from diagonalization routines are not necessarily exactly degenerate, but rather appear as near-degenerate. At the same time, eigenvectors corresponding to these near-degenerate levels are not uniquely defined, so an energy threshold must be introduced to identify the degeneracies.

\section{\label{sec:Results}Results and Discussion}

We calculated the electronic band structures of six TMDs (MoS$_2$, MoSe$_2$, MoTe$_2$, WS$_2$, WSe$_2$, WTe$_2$) in the 2H, 1T, and 1T' structural phases using our 4c all-electron GTO method. We compared our 2H and 1T results with the corresponding band structures presented in Ref.~\cite{miro2014atlas} obtained from the method of Te Velde and Baerends~\cite{tevelde1991precise}. For the 1T' phase, we used the \textsc{Vasp} program package~\cite{kresse1996efficient} to investigate how our all-electron method performs compared to the pseudopotential approximation. Due to the presence of heavy atoms, all TMDs considered exhibit significant scalar relativistic and SOC effects. We investigated three different levels of theory: nonrelativistic (nr) with infinite speed of light, Dirac-based 4c scalar relativistic (sr) without SOC, and the Dirac-based 4c fully relativistic (fr) with SOC included. In all our calculations, we employed the nonrelativistic GGA-type XC functional PBE~\cite{perdew1996generalized}. For the 4c calculations, the Gaussian finite nucleus model of Visscher and Dyall~\cite{visscher1997dirac} was used instead of the point nuclei in order to regularize the singularity of the lower components of the wave function evaluated at the atomic centers. All electronic structure optimizations were performed using the relaxed unit cell geometries taken from Refs.~\cite{miro2014atlas} (2H and 1T phases) and~\cite{qian2014quantum} (1T' phase). For the momentum-space Brillouin zone integrations, we used the $\Gamma$-centered mesh of $\vec{k}$ points of $11\times 11$ for the 2H and 1T' phases and $33\times 33$ for the metallic 1T phase. When comparing the band structures $\epsilon^{(1)}_n(\vec{k})$ and $\epsilon^{(2)}_n(\vec{k})$ obtained by using two different methods or computational settings, we evaluate the maximum difference between the energy eigenvalues as well as the root-mean-square deviation (``band delta'')~\cite{huhn2017one},
\begin{equation}
	\Delta_\text{b}(\mathcal{W}) = \sqrt{\frac{1}{N_E} \sum_{\substack{\vec{k},n\\ \epsilon^{(1)}_n(\vec{k}) \in\mathcal{W}\\ \epsilon^{(2)}_n(\vec{k}) \in\mathcal{W}}} \left( \epsilon^{(1)}_n(\vec{k}) - \epsilon^{(2)}_n(\vec{k}) \right)^2},
\label{eq:bandDelta}
\end{equation}
where $\mathcal{W} = [\epsilon_\text{l},\epsilon_\text{u}]$ is an energy window chosen for comparison, the summations run over all energy states along a path of $\vec{k}$ points for which both $\epsilon^{(1)}_n(\vec{k})$ and $\epsilon^{(2)}_n(\vec{k})$ lie inside the window $\mathcal{W}$, and $N_E$ is the total number of such states.

\subsection{\label{sec:Numerical}Numerical and basis-set accuracy}

Numerical instabilities in simulations involving periodic systems with GTOs originate from two distinct, albeit related, reasons~\cite{suhai1982error,kudin2000linear}. First, the numerical accuracy of the KS Hamiltonian matrix can be insufficient when various approximations necessary for solid-state simulations are introduced for basis sets with diffuse functions. Diffuse basis functions can extend over a very large number of unit cells, and the overlaps between such functions exhibit a much slower decay than their more localized counterparts. If these overlaps are neglected too early in calculations, the summations in Eqs.~\eqref{eq:kSpaceGroupFandS} are not converged, causing the overlap matrix $S_{\mu\mu'}(\vec{k})$ to be indefinite, \emph{i.e.} having negative eigenvalues. This problem can be mitigated by including a large number of basis function products $\dg{\chi}_{\mu}(\vec{r}) \chi_{\mu'}(\vec{r}-\vec{R})$, which in turn increases the memory requirements for storage of matrices. To this end, in the periodic module of \textsc{ReSpect}, we implemented sparse data structures in real space in combination with quaternion algebra to efficiently handle spin--orbit-coupled wave-function components. This versatile approach allows us to retain the accuracy necessary for numerical robustness with a very small additional memory footprint.

The second problem is associated with overcompleteness (linear dependencies) of the basis that causes matrices in Eq.~\eqref{eq:DKSinMatrixForm} to be ill-conditioned and introduces errors and instabilities in matrix operations. This occurs when the eigenvalues of the momentum-space overlap matrix drop below $10^{-7}$, which is often the case for larger basis sets. For instance, the smallest eigenvalue of the overlap matrix of WTe$_2$ in the 1T' phase is $10^{-11}$, while the remaining eigenvalues span over 12 orders of magnitude. As a result, very large elements $\sim 10^8$ accumulate in the density matrix, which prohibits the successful completion of the ground-state convergence procedure. To avoid such errors, we construct an orthonormal basis so that the redundant basis functions are projected out. However, we found that truncating the space spanned by orthonormal basis functions used for describing the lower wave-function components leads to the formation of artificial states inside the band gap for some $\vec{k}$ points in the band structure. This can be understood by realizing that the RKB condition is violated by the truncation process. To numerically control how much of the RKB is lost, we derived the matrix form of the RKB condition in a general basis and the expression for the truncation error. More details on the process of constructing the orthonormal basis in the 4c framework is described in Sec.~\ref{sec:Construction}.

For all atoms, we employed the relativistic triple-$\zeta$ (TZ) basis sets developed by Dyall for $d$~\cite{dyall2007relativistic,dyall2004relativistic,dyall2010revised} and $p$ elements~\cite{dyall2002relativistic,dyall2006relativistic}. The basis sets used were uncontracted and contained high-angular momentum correlating functions for several outer shells and functions for dipole polarization of the valence shells. Our initial tests using the smaller double-$\zeta$ (DZ) basis from the same family yielded results with inconsistent quality across $\vec{k}$ points -- for instance, the band energy at the $M$ point of 2H MoS$_2$ was improved by 20 meV when the TZ basis was used, while for other $\vec{k}$ points, this difference was only about 6 meV. Similarly, the differences in the band structures of 2H WS$_2$ (including the band gap) between the DZ and TZ basis sets were 20-30 meV. Hence, to ensure that our results are well-converged for all $\vec{k}$ points and that the target accuracy is sufficient for reliable comparisons with other methods, we chose the TZ basis in all calculations reported here.

The basis sets employed contained very diffuse functions with several exponents of $<10^{-2}$ a.u. The extent of the most diffuse function on Mo and W was 18.2 and 17.2 \AA, respectively, causing the basis function to span more than a hundred unit cells of the 2D film. This in turn results in several hundred thousand charge distributions interacting with themselves and their periodic replicas when the four-center Coulomb integrals are evaluated. Such calculations are currently unfeasible on common supercomputers as they can consume a notable portion of allocated resources. However, within the RI approximation, the charge distributions only interact with the electron density that is expanded using an atom-centered auxiliary basis. Since the number of the three-center terms needed is smaller than the number of four-center terms typically by three orders of magnitude, the evaluation of the Coulomb operator becomes affordable. Tests that we performed on various systems including selected 2H and 1T' structures studied here in both the nr and fr settings indicate that the maximum error introduced by the RI approximation is very small---only a few tens of $\mu$eV.

Our implementation allowed us to explore the validity of removing diffuse functions from the GTO basis sets before starting the electronic-structure optimization --- an approach that is commonly employed for accelerating and numerically stabilizing calculations of periodic systems using GTOs~\cite{peralta2005scalar,usvyat2011approaching,perry2001antiferromagnetic}. From the TZ basis sets, we deleted all exponents $<0.1$ which for most atoms meant removing one to two functions in each of the $s$, $p$, $d$, and $f$ shells. Even though this modification reduced the computational cost by a factor between five and six in case of the nr theory and a factor of four in case of the 4c theory, we observed lowered accuracy of the eigenenergies of some bands and $\vec{k}$ points. For instance, the difference between the energy of the $\Gamma$ point of the valence band of 2H WTe$_2$ obtained from the calculations using the truncated and full basis sets was 0.1 eV. The virtual bands located 2.5 eV above the Fermi level were, in general, poorly described with the truncated basis set. Hence, our results indicate that removing the diffuse functions in the start of the calculation is neither justified nor necessary.

\subsection{\label{sec:Phases}Stability of phases}

\begin{figure}[tb]
\includegraphics[width=1.00\columnwidth]{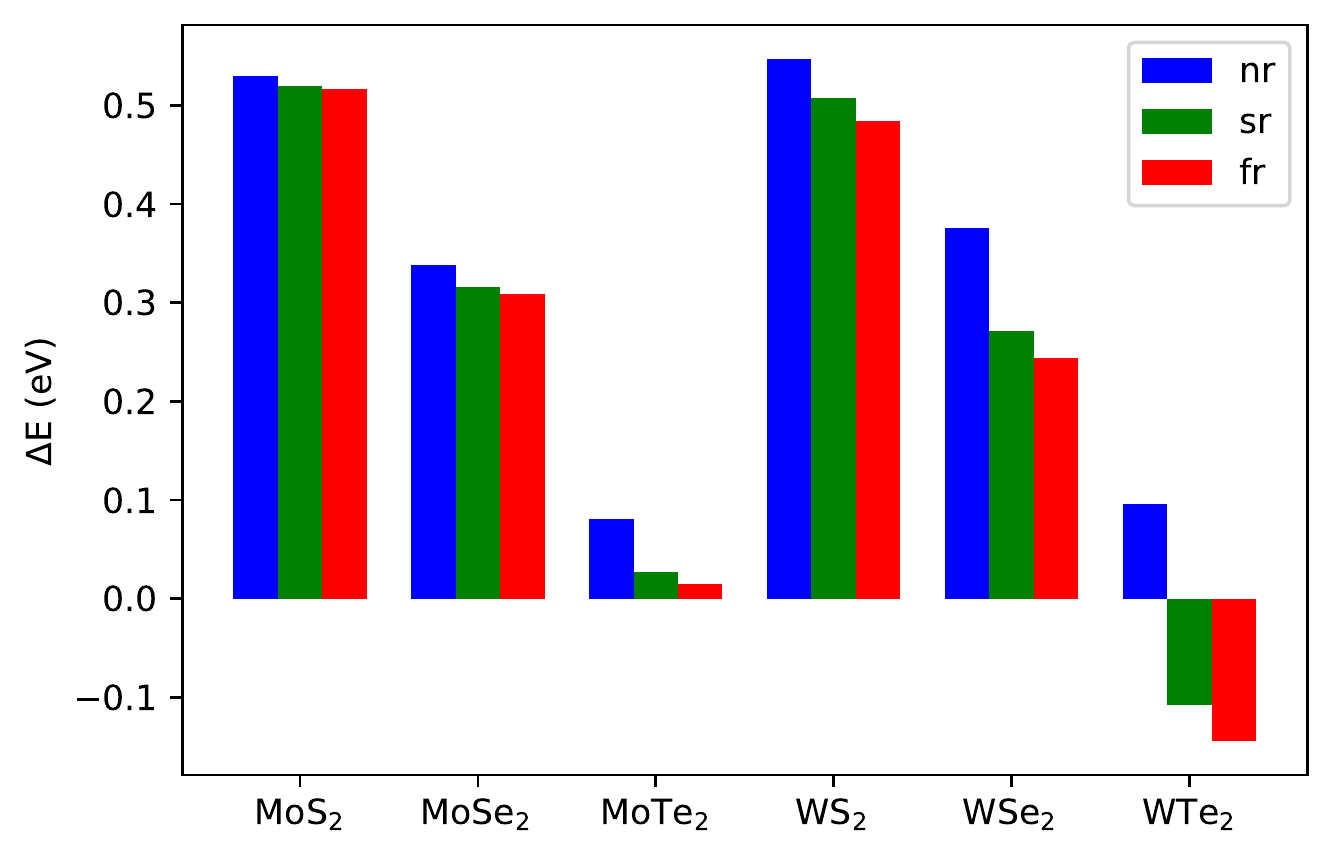}
\caption{\label{fig:phases} Relative total energy of the 1T' phase of TMD monolayers per $MX_2$ calculated with respect to the 2H phase. The results were obtained for nr, sr, and fr Hamiltonians and show the relativistic effects on the structural stability.}
\end{figure}

Before inspecting individual structural phases and validating our approach, we compare the total energies of each phase. Our results confirm the findings of Qian \emph{et al.}~\cite{qian2014quantum} --- for all systems except WTe$_2$, the 2H phase was found to be the ground-state structure with the lowest energy, followed by the higher-energy 1T' phase. The metallic 1T phase is thermodynamically unstable and the systems undergo Jahn--Teller distortion, relaxing to the 1T' phase~\cite{qian2014quantum}. In the case of WTe$_2$, the energy of the 1T' phase is below the 2H phase, which makes 1T' the most stable phase.

Our study allows for a uniform treatment of relativistic effects for all atoms and all electrons, as opposed to the calculations employing pseudopotentials, which offer less control over how relativity is handled for light and heavy elements. We found that it is the inclusion of the scalar relativistic effects that decreases the energy of the 1T' phase with respect to the 2H phase. Including the SOC further lowers the energy of the 1T' phase. For the heaviest system, WTe$_2$, this energy lowering becomes sufficient to change the ground-state phase from 2H to 1T'. Figure~\ref{fig:phases} summarizes our results for all systems at the nr, sr, and fr levels of theory and shows the relative energy differences between the 1T' and 2H phases. Hence, without relativistic effects, the ground-state structure of all six systems would be 2H.

\subsection{\label{sec:Results2H1T}2H and 1T phases}

\begin{table}[tb]
\caption{\label{tab:gaps1H} Band gaps and SOC-induced (Rashba) splitting of the valence band at the $K$ point of TMD monolayers in the 2H phase calculated for nr, sr, and fr Hamiltonians. The values in the round and square brackets are taken from Refs.~\cite{miro2014atlas} and~\cite{kosmider2013large}, respectively.}
\begin{ruledtabular}
\begin{tabular}{cdddd}
& \multicolumn{3}{c}{Band gap (E$_g$, eV)} & \multicolumn{1}{c}{Rashba splitting (eV)} \\
\cline{2-5}
& \multicolumn{1}{c}{nr} & \multicolumn{1}{c}{sr} & \multicolumn{1}{c}{fr} & \multicolumn{1}{c}{fr} \\
\hline
MoS$_2$ &  1.820  &  1.812  &  1.734   &   0.146  \\
        &         & (1.82)  & (1.74)   &  [0.147] \\
MoSe$_2$&  1.560  &  1.550  &  1.442   &   0.185  \\
        &         & (1.56)  & (1.45)   &  [0.186] \\
MoTe$_2$&  1.171  &  1.154  &  1.018   &   0.218  \\
        &         & (1.15)  & (1.01)   &          \\
WS$_2$  &  2.060  &  1.984  &  1.628   &   0.416  \\
        &         & (1.98)  & (1.64)   &  [0.433] \\
WSe$_2$ &  1.741  &  1.630  &  1.300   &   0.460  \\
        &         & (1.63)  & (1.33)   &  [0.463] \\
WTe$_2$ &  1.332  &  1.194  &  0.863   &   0.479  \\
        &         & (1.18)  & (0.87)   &
\end{tabular}
\end{ruledtabular}
\end{table}

The 2H and 1T phases of the TMDs both contain three atoms (one metal and two chalcogens) in the primitive unit cell. However, the structural differences between the two phases lead to their distinct electronic properties. Whereas the 1T phase is space-inversion-symmetric with a metallic electronic structure, the 2H systems are semiconductors with broken space inversion. In combination with the strong SOC, this broken inversion symmetry leads to ``giant'' spin--orbit-induced Rashba splittings in 2H $MX_2$~\cite{zhu2011giant}.

To validate the approach presented in this work, we calculated the band structures at the nr, sr, and fr levels of theory using all-electron bases. The resulting band diagrams traversing high symmetry $\vec{k}$ points obtained with and without SOC can be found in Figs.~S1 (2H) and~S2 (1T) in Ref.~\cite{sm}. The nonrelativistic results differ significantly from those obtained from the relativistic theories and are not shown. For the 2H and 1T phases, sr band structures are available in the 2D materials Atlas of Mir\'o, Audiffred, and Heine~\cite{miro2014atlas} that we used for comparison. We conclude that our results agree very well at the sr level with the Atlas results.

In the MoTe$_2$ and WTe$_2$ band structures of the 1T phase shown in Fig.~S2~\cite{sm}, it is possible to see very small SOC-induced ``Rashba-like'' spin splittings of the order of tens of meV. However, such splittings should not be present for the inversion-symmetric 1T phase, and all the bands should be strictly doubly degenerate. We found that this observation can be attributed to small deviations (about 0.04 \AA) in the unit cell geometry of MoTe$_2$ and WTe$_2$ from the tetragonal lattice, that is, the 1T structure used does not exhibit exact inversion symmetry. The geometries of the other systems do not show any deviations from 1T, and we observed exact double degeneracy of bands. Finally, we note that including the SOC in the computational framework lifted the degeneracy at the intersection points of the overlapping valence and conduction bands of the metallic 1T phase. The values of the SOC-induced vertical gaps at the $K$ point are reported in Table~SI of Ref.~\cite{sm}.

Table~\ref{tab:gaps1H} contains the values of the nr, sr, and fr band gaps of the 2H phase, as well as the Rashba splittings of the valence band at the $K$ point. We attribute the small discrepancies between the results obtained with the method presented here and the Atlas approach to differences in the methodologies. In the Atlas work, the relativistic effects are treated with the zeroth-order regular approximation (ZORA)~\cite{van1996relativistic}, which is a 2c technique that requires numerical integration schemes due to the appearance of the potential in the denominator of the Hamiltonian. In contrast, the Hamiltonian used here is 4c with most of the integrals evaluated analytically. The basis sets employed should not cause any significant deviations, since both sets of calculations were carried out with TZ quality bases with several polarization functions. Finally, we note that the Atlas results included empirical D3 treatment of London dispersion interactions~\cite{grimme2010consistent} that we did not consider in this work. The fact that no larger differences between the two approaches were observed even when the dispersion corrections were omitted indicates that the band gaps in thin monolayers are dominated by covalent bonds, which are well described without these corrections. However, the dispersion forces are important when multiple layers of 2D materials are held together by van der Waals forces as, for example, in bilayers and heterostructures~\cite{liu2023role}. In the case of WSe$_2$, for which we observed the largest difference of 30 meV in the fr setting, we further benchmarked against a calculation with a larger quadruple-$\zeta$ (QZ) basis set. The band energies improved only marginally, \emph{i.e.} by units of meV. Hence, we believe that our band structures are well converged with respect to the basis set as well as the relativistic Hamiltonian and should serve as the reference results for these systems for the PBE functional.

\subsection{\label{sec:Results1Tp}1T' phase}

% \begin{table}[tb]
% \caption{\label{tab:gaps1Tp} 1T' phase. The values in the square brackets are our \textsc{Vasp} calculations. The values in the round brackets are taken from Ref.~\cite{qian2014quantum}. \fixme{caption}}
% \begin{ruledtabular}
% \begin{tabular}{ccccc}
% & \multicolumn{2}{c}{Fundamental gap (E$_g$, meV)} & \multicolumn{2}{c}{Inverted gap (2$\delta$, meV)} \\
% \cline{2-5}
% & \multicolumn{1}{c}{DFT-PBE} & \multicolumn{1}{c}{$G_0W_0$} & \multicolumn{1}{c}{DFT-PBE} & \multicolumn{1}{c}{$G_0W_0$} \\
% \hline
% MoS$_2$  & 48 && 477 & \\
%          &  [47] && [536] & \\
%          &  (45) &  (76)  & (540) & (562) \\
% MoSe$_2$ & 32 && 734 & \\
%          & [32] && [715] & \\
%          &  (31) &  (88)  & (706) & (988) \\
% MoTe$_2$ & -242 && 374 & \\
%          & [-260] && [345] & \\
%          & (-262) & (-300)  & (344) & (403) \\
% WS$_2$   & 14 && 136 & \\
% %WS$_2$   & 72 && 136 & \\
%          & [85] && [197] & \\
%          &  (44) &  (110)  & (187) & (284) \\
% WSe$_2$  & 27 && 660 & \\
%          & [35] && [694] & \\
%          &  (36) &  (116)  & (701) & (863) \\
% WTe$_2$  & -102 && 974 & \\
%          & [-103] && [961] & \\
%          & (-112) & (-133)  & (952) & (978)
% \end{tabular}
% \end{ruledtabular}
% \end{table}

\begin{figure}[tb]
\includegraphics[width=0.99\columnwidth]{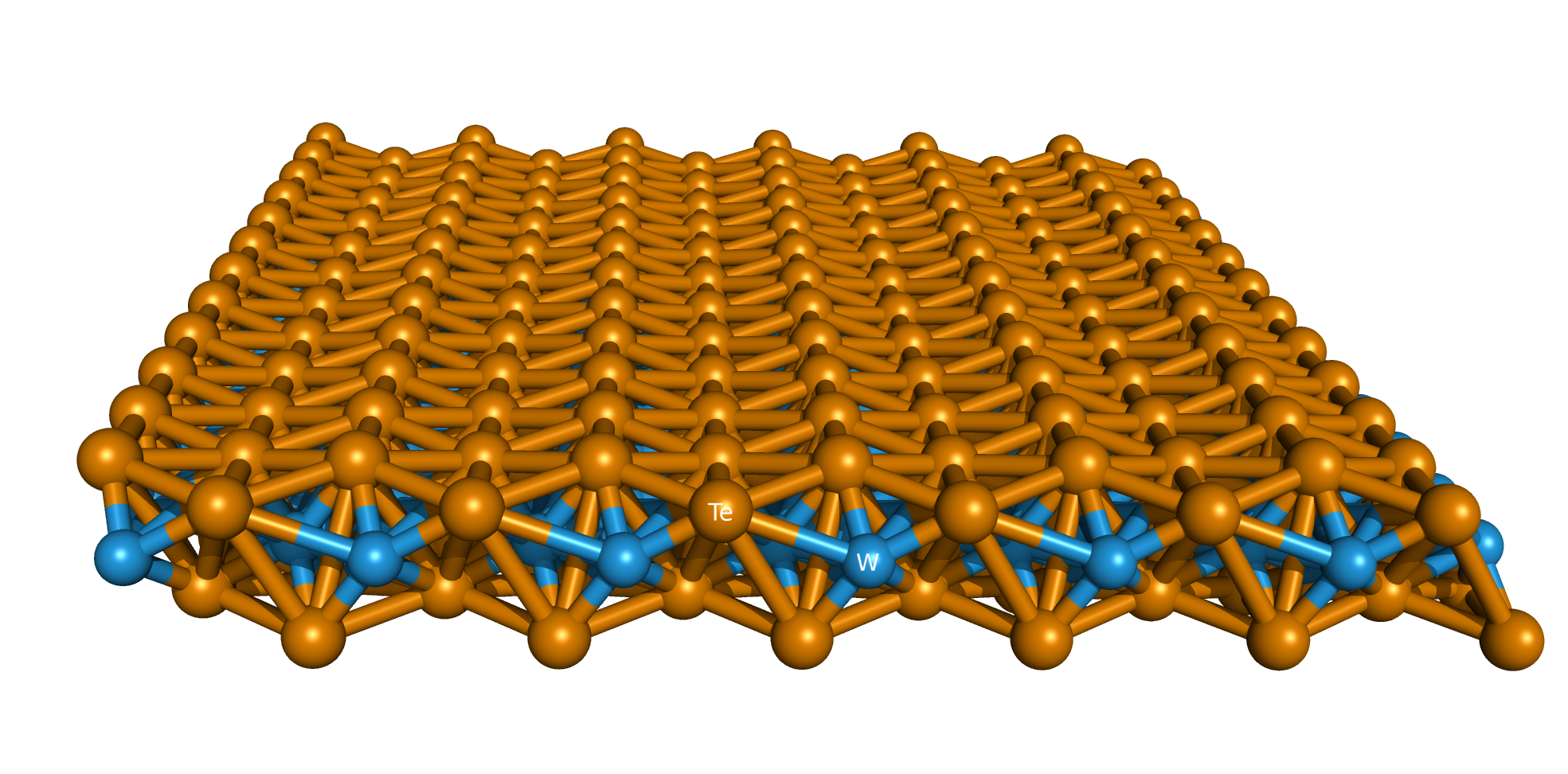}
\caption{\label{fig:mx2struct} Structure of the 1T' phase of monolayer TMDs.}
\end{figure}

\begin{figure*}[tb]
\includegraphics[width=0.32\textwidth]{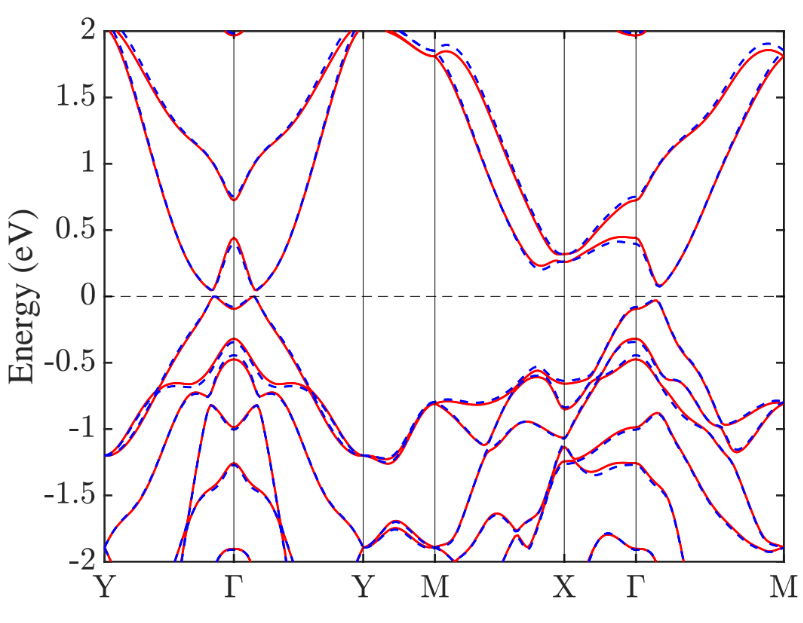}
\includegraphics[width=0.32\textwidth]{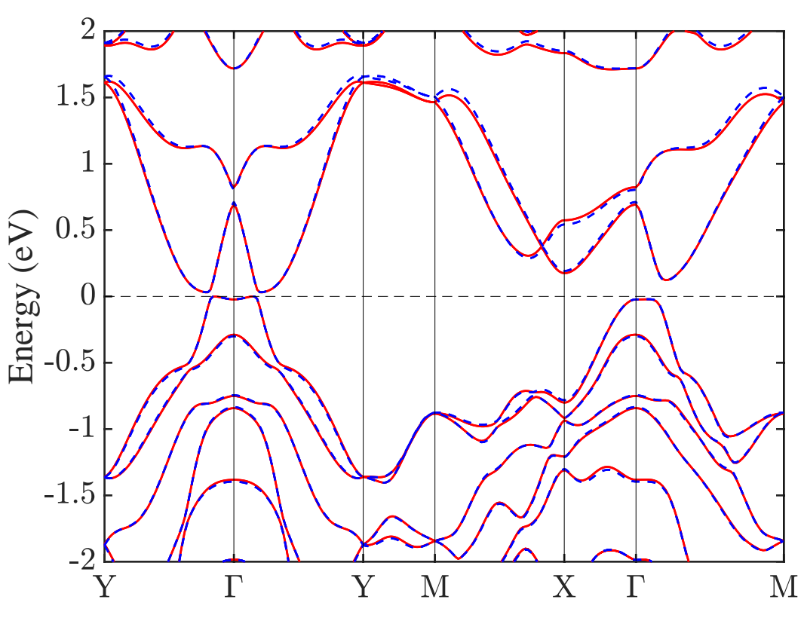}
\includegraphics[width=0.32\textwidth]{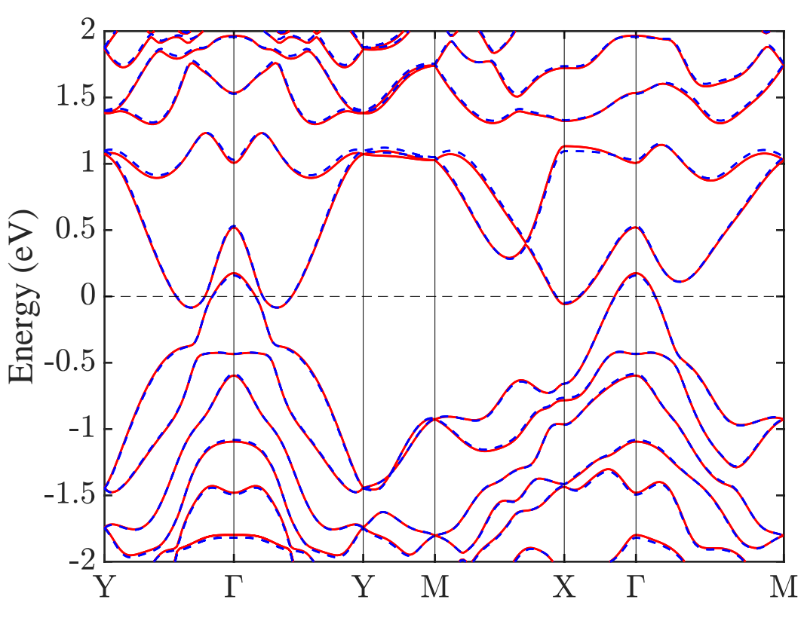}
\includegraphics[width=0.32\textwidth]{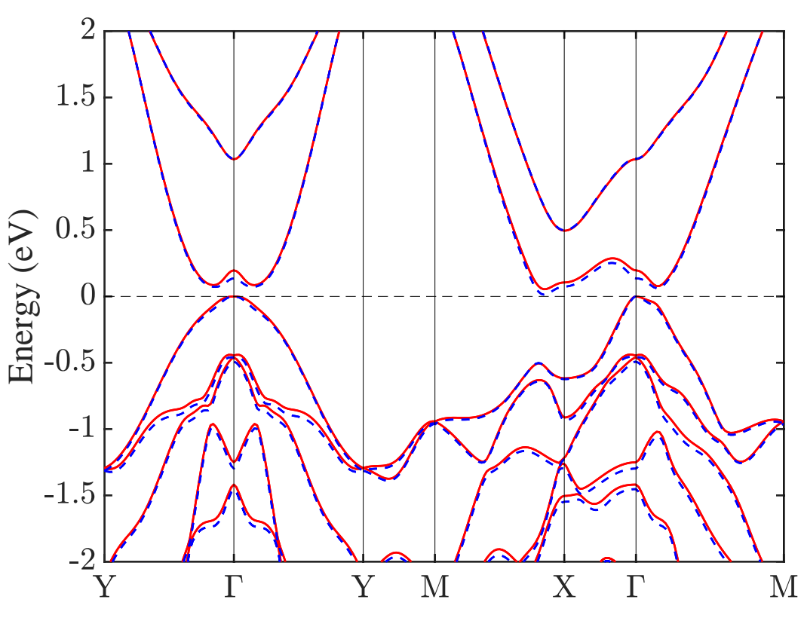}
\includegraphics[width=0.32\textwidth]{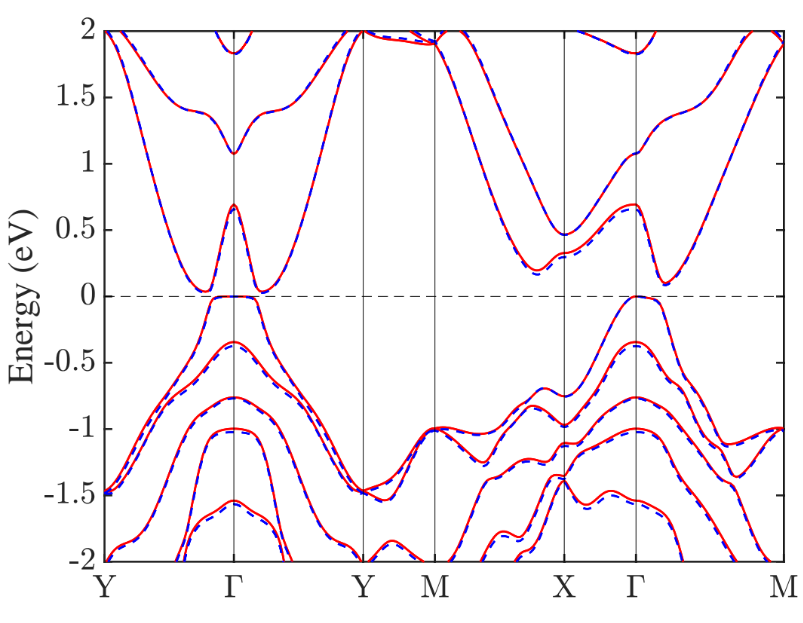}
\includegraphics[width=0.32\textwidth]{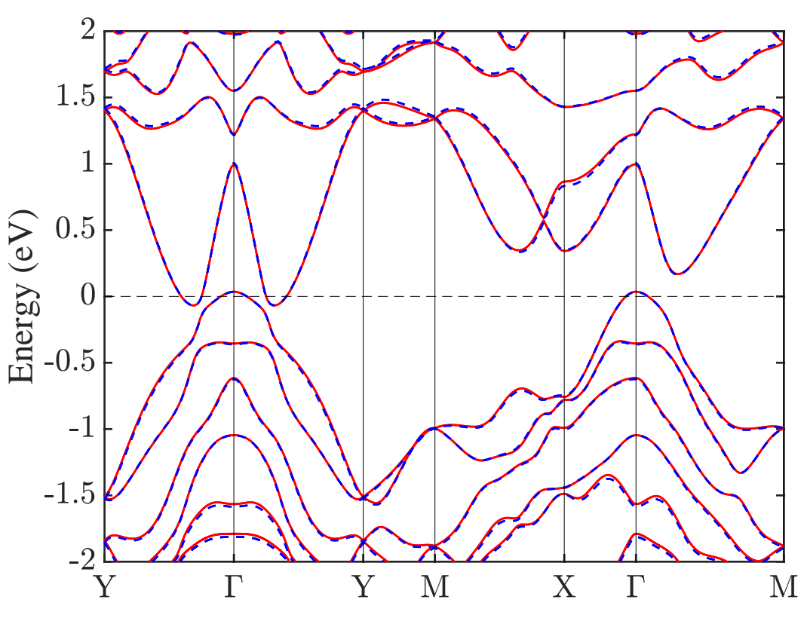}
\caption{\label{fig:mx2bsTetPVasp} Comparison of band structures of $MX_2$ monolayers in the 1T' phase obtained from the \textsc{ReSpect} (dashed lines) and \textsc{Vasp} (full lines) codes at the relativistic level of theory including SOC. From the top left to bottom right: MoS$_2$, MoSe$_2$, MoTe$_2$, WS$_2$, WSe$_2$, and WTe$_2$. The horizontal dashed black line marks the Fermi level. The path traversing high-symmetry $\vec{k}$-points in the reciprocal-space unit cell was chosen according to Ref.~\cite{setyawan2010high}.}
\end{figure*}

\begin{table}[tb]
\caption{\label{tab:delta1Tp} Root-mean-square deviation between the band structures of TMD monolayers obtained from the \textsc{ReSpect} and \textsc{Vasp} codes at the fr level of theory (including SOC) using Eq.~\eqref{eq:bandDelta}. The valence region $\mathcal{W}_\text{val}$ ranges from $-10$ to 0 eV (shifted to the Fermi level) and the conduction region $\mathcal{W}_\text{con}$ ranges from 0 to 5 eV.}
\begin{ruledtabular}
\begin{tabular}{cdddddd}
\multicolumn{7}{c}{Band delta (meV)}\\
  & \multicolumn{1}{c}{MoS$_2$} & \multicolumn{1}{c}{MoSe$_2$} & \multicolumn{1}{c}{MoTe$_2$} & \multicolumn{1}{c}{WS$_2$} & \multicolumn{1}{c}{WSe$_2$} & \multicolumn{1}{c}{WTe$_2$} \\
\hline
$\Delta_\text{b}(\mathcal{W}_\text{val})$ & 12.0 &    8.4 & 11.7 & 34.6 & 25.6 & 20.2 \\
$\Delta_\text{b}(\mathcal{W}_\text{con})$ & 40.5 &   28.4 & 32.8 & 26.2 & 23.4 & 42.0
\end{tabular}
\end{ruledtabular}
\end{table}

The 1T' phase of $MX_2$ is formed from the 1T structure by a spontaneous lattice distortion in which the unit cell period is doubled along one in-plane direction, which creates zigzag chains along the perpendicular in-plane direction~\cite{eda2012coherent} (see Fig.~\ref{fig:mx2struct}). The TMDs were shown to host the QSH effect in the 1T' structure, \emph{i.e.} the existence of time-reversal-symmetry protected edge states that enable conduction of electrical currents on the surface of the material~\cite{qian2014quantum}. From a computational viewpoint, the 1T' phase has not been studied in the literature to the same extent as the 2H and 1T counterparts. Here, we calculated the electronic band structures of the 1T' $MX_2$ using the all-electron 4c method presented in this work as well as the pseudopotential method as implemented in the \textsc{Vasp} package~\cite{kresse1996efficient}. For the \textsc{Vasp} calculations, we used an energy cutoff of 500 eV and the Brillouin zone $\vec{k}$-point sampling of $11\times 15\times 1$. We also tested a denser mesh of $\vec{k}$ points without observing significant differences. To avoid artificial interactions between the periodic images of the $MX_2$ layers that appear when 2D systems are studied using plane waves with periodic boundary conditions, a vacuum region of 20 \AA{} was applied in the direction perpendicular to the 2D film.

The band structures obtained from the \textsc{ReSpect} and \textsc{Vasp} codes with SOC are presented in Fig.~\ref{fig:mx2bsTetPVasp}. The values of the band delta differences between the two codes are shown in Table~\ref{tab:delta1Tp}, evaluated separately for the occupied (valence) and virtual (conduction) energy regions. In general, the differences are of the order of tens of meV and are larger in the virtual region. The biggest discrepancies close to the Fermi level are seen in WS$_2$, while the overall largest band delta values are found in MoS$_2$ and WTe$_2$. For comparison, we performed tests of the numerical convergence of the GTO method by tuning various parameters of the \textsc{ReSpect} code. In particular, we modified the number of removed momentum-space basis functions in the orthonormal basis (see Sec.~\ref{sec:Construction}), tightened the lattice-sum convergence thresholds, as well as turned off the RI approximation of the Coulomb integrals. Such modifications had a negligible effect on our GTO results; the band delta values between the higher accuracy calculations and the results shown here reached tens of $\mu$eV. Bigger differences were observed only when we benchmarked the WS$_2$ TZ calculation against the QZ basis set. Still, the energies of the bands close to the Fermi level changed by 1--6 meV, and the values of the band deltas for the intervals $-10$ to $0$ eV and $0$ to $5$ eV were $\Delta_\text{b}(\mathcal{W}_\text{val}) \approx 4$ meV and $\Delta_\text{b}(\mathcal{W}_\text{con}) \approx 10$ meV, respectively. To conclude, despite the minor improvements to the band structure that are possible when larger basis sets are employed, our TZ results can be considered as the reference for benchmarking performance of other approximate methods, for instance, for studying the accuracy of various pseudopotentials.

We explored the topological nature of the 1T' phase of the $MX_2$ systems by first verifying that the band-gap opening is due to the SOC. Fig.~S3 in Ref.~\cite{sm} shows that energy degeneracies occurring at points where the valence band crosses the conduction band are lifted when SOC is included. A band gap opens in all systems except MoTe$_2$ and WTe$_2$ that remain metallic despite the band inversion. The lattice distortion in the 1T' phase of the 2D TMDs does not break the inversion symmetry. Hence, we used the parity of the Bloch states evaluated at the TRIM points to calculate the $\mathbb{Z}_2$ invariant~\cite{fu2007topological}. Our results for the parities of the individual Kramers pairs agree with those obtained by Qian \email{et al.}~\cite{qian2014quantum}, which yields $(-1)^{\nu} = 1$ for all six TMDs and confirms the existence of the surface edge states hallmarking the quantum spin Hall phase.

\section{\label{sec:Conclusion}Conclusion and outlook}

2D TMDs are attracting high current interest for hosting exotic phenomena that promise applications in spintronic devices and quantum computing. Theoretical predictions and understanding of such applications relies on accurate description of the relativistic effects driven by the presence of heavy atoms in these materials. Accordingly, we have presented here an all-electron approach that builds on the most widely used basis sets in quantum chemistry to obtain a scheme that can accurately capture relativistic effects as well as the properties of electron wave functions close to the nuclei within a uniform theoretical framework applicable to both solids and molecules.

GTOs provide a convenient link between the finite and periodic or extended systems, yet their use for solids has been met with many technical difficulties that sparked skepticism in the community; it has even been suggested that GTOs should be avoided altogether for condensed matter calculations~\cite{jensen2013analysis}. However, in this work, we show how the common limitations associated with GTOs can be overcome, and that the convergence with respect to the basis-set limit is possible even in the fully relativistic 4c setting. To mitigate numerical and performance issues, our method is based on the quaternion algebra, linear-scaling data structures, RI approximation of the Coulomb integrals, orthonormal momentum-space bases that sufficiently preserve the RKB condition, and a parallel electronic structure solver with minimal required IO operations. Our in-depth analysis of the 2H, 1T, and 1T' phases of 2D TMDs indicates that our all-electron 4c results can serve as the reference for developing pseudopotentials or approximate 2c relativistic techniques. We confirm the existence of the quantum spin Hall effect in the 1T' phase of $MX_2$ by evaluating the $\mathbb{Z}_2$ invariant within our real-space formulation. Finally, we attribute the origin of the lower ground-state energy of WTe$_2$ in the 1T' compared to the 2H phase to (scalar) relativistic effects.

Despite the good agreement between our band structures for the 2H and 1T phases and those obtained by using the 2c ZORA method~\cite{miro2014atlas}, our 4c approach offers several advantages. The exact Dirac Hamiltonian contains the mass-velocity term that is missing in the ZORA approximation~\cite{saue2011relativistic}, but it is important for an accurate description of relativistic effects in heavy elements and for evaluating properties in the presence of external magnetic fields~\cite{visscher1999full}. Also, the 4c DKS equation does not contain the potential in the denominator, so that complicated numerical integration of most integrals involved can be avoided.

Our study is important for opening a pathway for further developments that leverage quantum chemistry methods for exploring properties that are strongly affected by the relativistic effects without sacrificing access to the core regions of the electron charge and spin densities. For instance, accurate relativistic first-principles calculations of the spin Hamiltonian parameters (\emph{e.g.} hyperfine coupling constants, Zeeman interactions, or zero-field splittings) are needed for determining spin dynamical properties of spin defects in semiconductors~\cite{ghosh2021spin} to help find viable qubit hosting materials. Our all-electron method can also be used for generating accurate pseudopotentials, particularly for systems that contain heavy elements in high oxidation states.

% \section*{\label{sec:Data}Data availability}
% The data presented in this study as well as the scripts used for pre- and post-processing of the input and output files are available in the \href{https://doi.org/10.5281/zenodo.7394905}{ZENODO public repository}~\cite{dataset}.

% \section*{\label{sec:Code}Code availability}
% The \textsc{ReSpect} code used in this study is available at \href{www.respectprogram.org}{www.respectprogram.org} upon a reasonable request free of charge.

\begin{acknowledgments}
This work was supported by the Research Council of Norway through its Centres of Excellence scheme (Grant No.~262695), a Research Grant (Grant No.~315822), and its Mobility Grant scheme (Grant No.~301864), as well as the use of computational resources provided by UNINETT Sigma2 -- The National Infrastructure for High Performance Computing and Data Storage in Norway (Grant No. NN4654K). In addition, this project received funding from the European Union's Horizon 2020 research and innovation program under the Marie Sk\l{}odowska-Curie Grant Agreement No.~945478 (SASPRO2), and the Slovak Research and Development Agency (Grant No.~APVV-21-0497). The work at Northeastern University was supported by the US Department of Energy (DOE), Office of Science, Basic Energy Sciences Grant No. DE-SC0022216 and benefited from Northeastern University’s Advanced Scientific Computation Center and the Discovery Cluster and the National Energy Research Scientific Computing Center through DOE Grant No. DE-AC02-05CH11231.
\end{acknowledgments}

\bibliography{references}% Produces the bibliography via BibTeX.

% \section*{\label{sec:Contributions}Author contributions}
% M.K. conceived the project, developed the electronic structure solver, parallelized all time-consuming parts of the code, calculated the final band structures of 2H, 1T, and 1T' phases using \textsc{ReSpect}, performed extensive numerical tests and analysis of the results. M.J. developed the theory and implementation of RI approximation in the relativistic setting. B.W. suggested studying the 1T' phase and calculated the band structures of 2H and 1T' phases using \textsc{Vasp}. B.W., W.C., and M.K. discussed SOC-driven properties of TMD monolayers, and evaluation of topological invariants. F.M. performed initial tests, analysis, and benchmark calculations of the pilot implementation of \textsc{ReSpect} applied on the ``Atlas'' data set of 2D materials and a study of TMD monolayers in 2H and 1T phases comparing with other codes based on atom-centered orbitals. M.R. provided guidance and supervision on the development of the RI approximation. K.R. and A.B. supervised and discussed the project. M.K. wrote the initial version of the entire manuscript, A.B. and K.R. edited the manuscript, and all authors reviewed and commented on the manuscript.

% \section*{\label{sec:Interests}Competing interests}
% The authors declare no competing financial or non-financial interests.

% \section*{\label{sec:Additional}Additional information}
% \textbf{Correspondence} and requests for materials should be addressed to Marius Kadek.

\end{document}

%\preprint{APS/123-QED}

\title{Band structures and $\mathbb{Z}_2$ invariants of two-dimensional transition metal dichalcogenide monolayers from fully-relativistic Dirac--Kohn--Sham theory using Gaussian-type orbitals}

\author{Marius Kadek}
\email{marius.kadek@uit.no}
\affiliation{Department of Physics, Northeastern University, Boston, Massachusetts 02115, USA}
\affiliation{Hylleraas Centre for Quantum Molecular Sciences, Department of Chemistry, UiT The Arctic University of Norway, N-9037 Troms\o, Norway}

\author{Baokai Wang}
\affiliation{Department of Physics, Northeastern University, Boston, Massachusetts 02115, USA}

\author{Marc Joosten}
\affiliation{Hylleraas Centre for Quantum Molecular Sciences, Department of Chemistry, UiT The Arctic University of Norway, N-9037 Troms\o, Norway}

\author{Wei-Chi Chiu}
\affiliation{Department of Physics, Northeastern University, Boston, Massachusetts 02115, USA}

\author{Francois Mairesse}   %Francois.mairesse@student.unamur.be
\affiliation{Laboratory of Theoretical Chemistry, University of Namur, B-5000 Namur, Belgium}

\author{Michal Repisky}
\affiliation{Hylleraas Centre for Quantum Molecular Sciences, Department of Chemistry, UiT The Arctic University of Norway, N-9037 Troms\o, Norway}
\affiliation{Department of Physical and Theoretical Chemistry, Faculty of Natural Sciences, Comenius University, Bratislava, Slovakia}

\author{Kenneth Ruud}
\affiliation{Hylleraas Centre for Quantum Molecular Sciences, Department of Chemistry, UiT The Arctic University of Norway, N-9037 Troms\o, Norway}
\affiliation{Norwegian Defence Research Establishment, P.O. Box 25, 2027 Kjeller, Norway}

\author{Arun Bansil}
\affiliation{Department of Physics, Northeastern University, Boston, Massachusetts 02115, USA}


\date{\today}% It is always \today, today,

\begin{center}
    \LARGE Supplemental Material for
\end{center}

\maketitle

%\tableofcontents

\begin{figure*}[tb]
\includegraphics[width=0.32\textwidth]{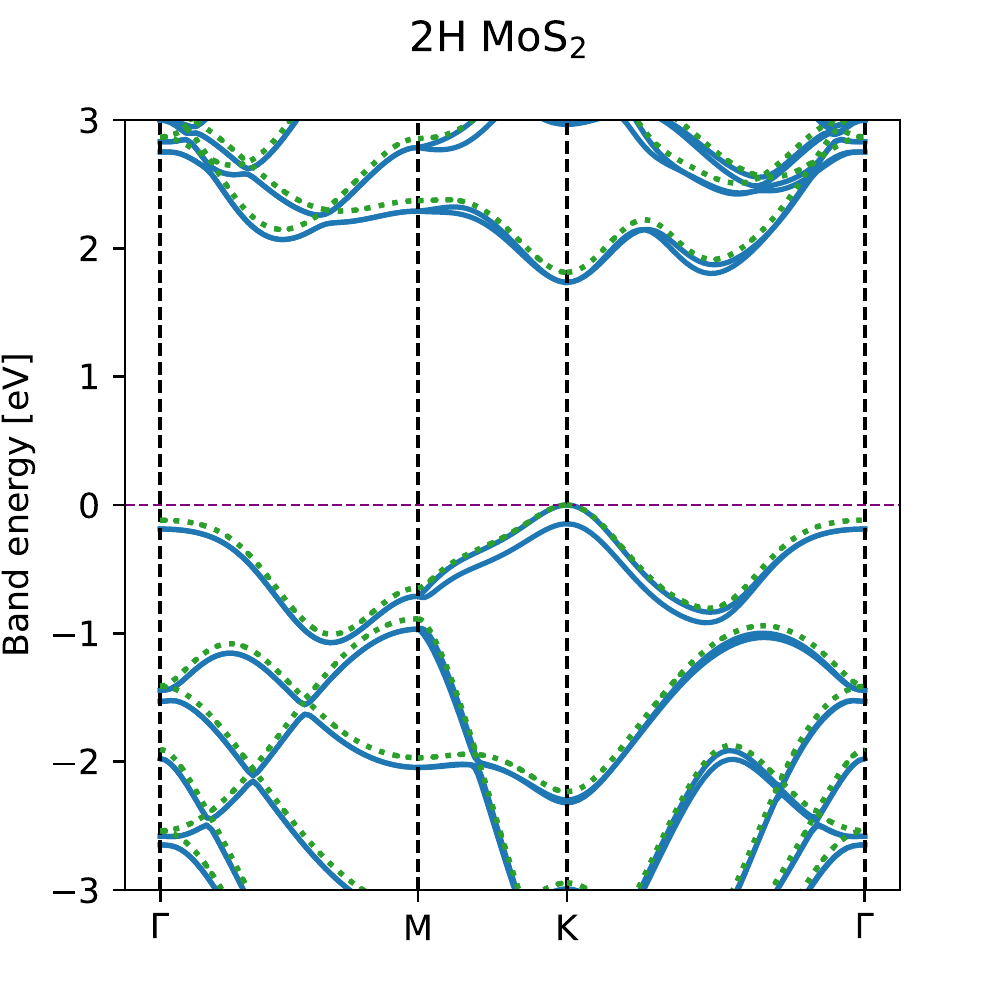}
\includegraphics[width=0.32\textwidth]{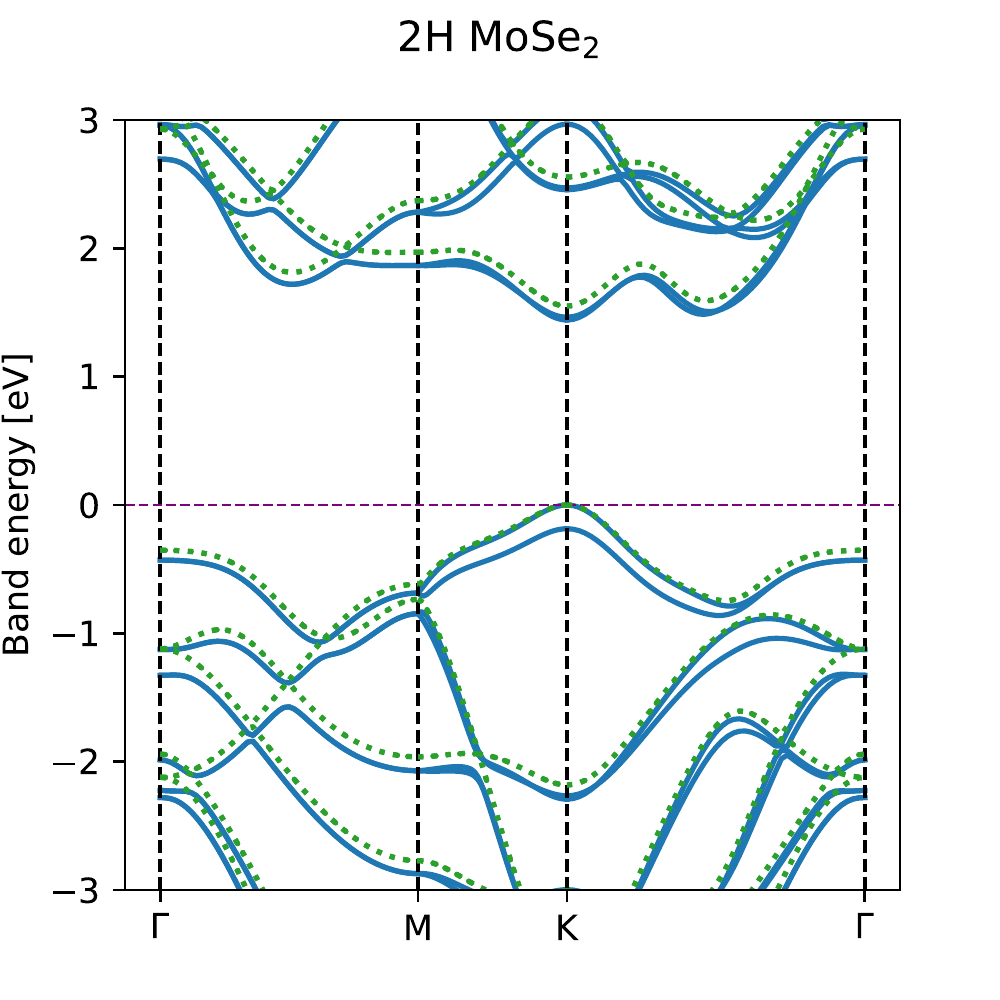}
\includegraphics[width=0.32\textwidth]{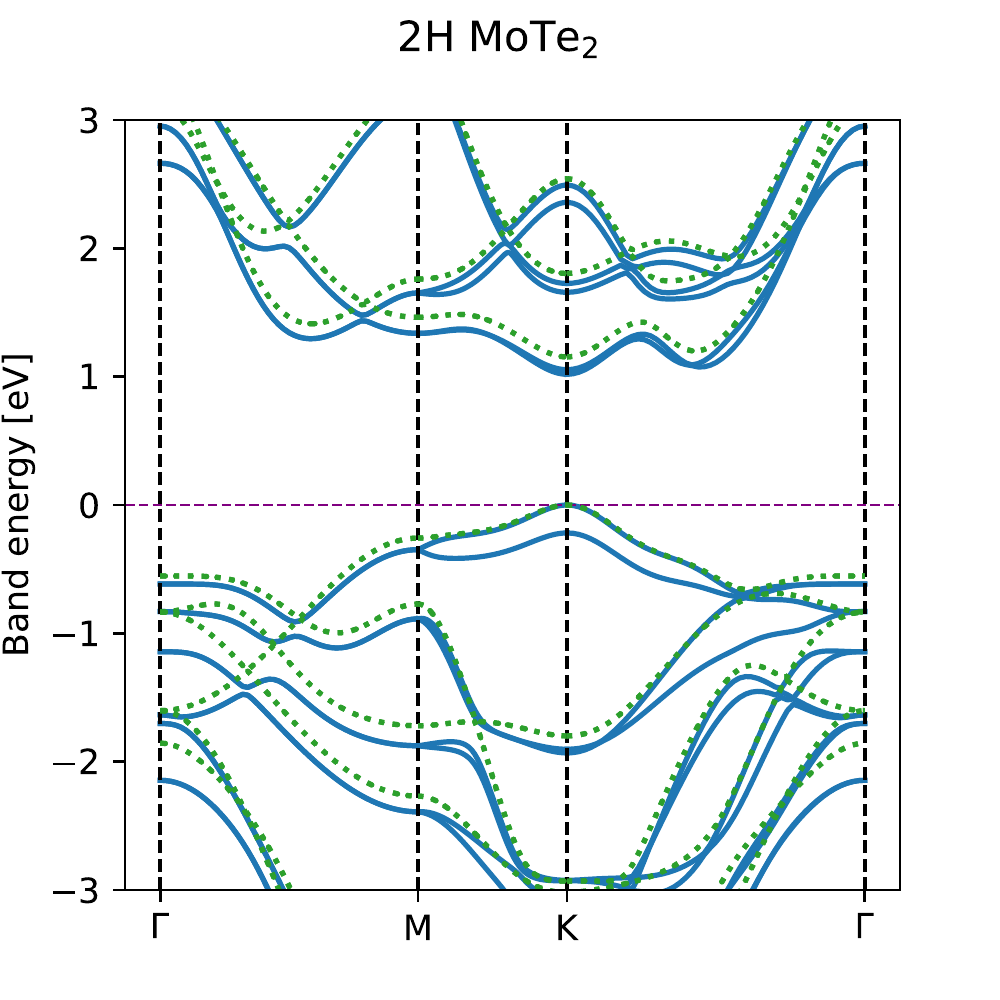}
\includegraphics[width=0.32\textwidth]{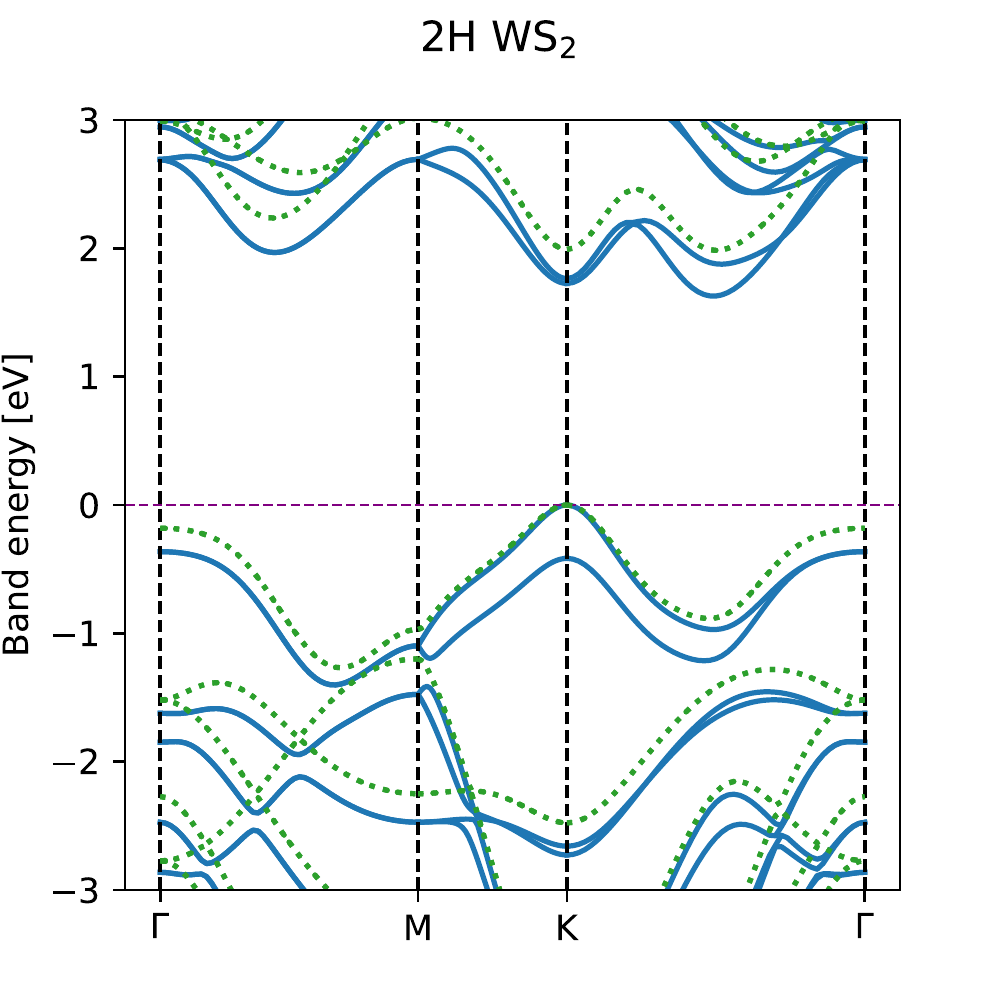}
\includegraphics[width=0.32\textwidth]{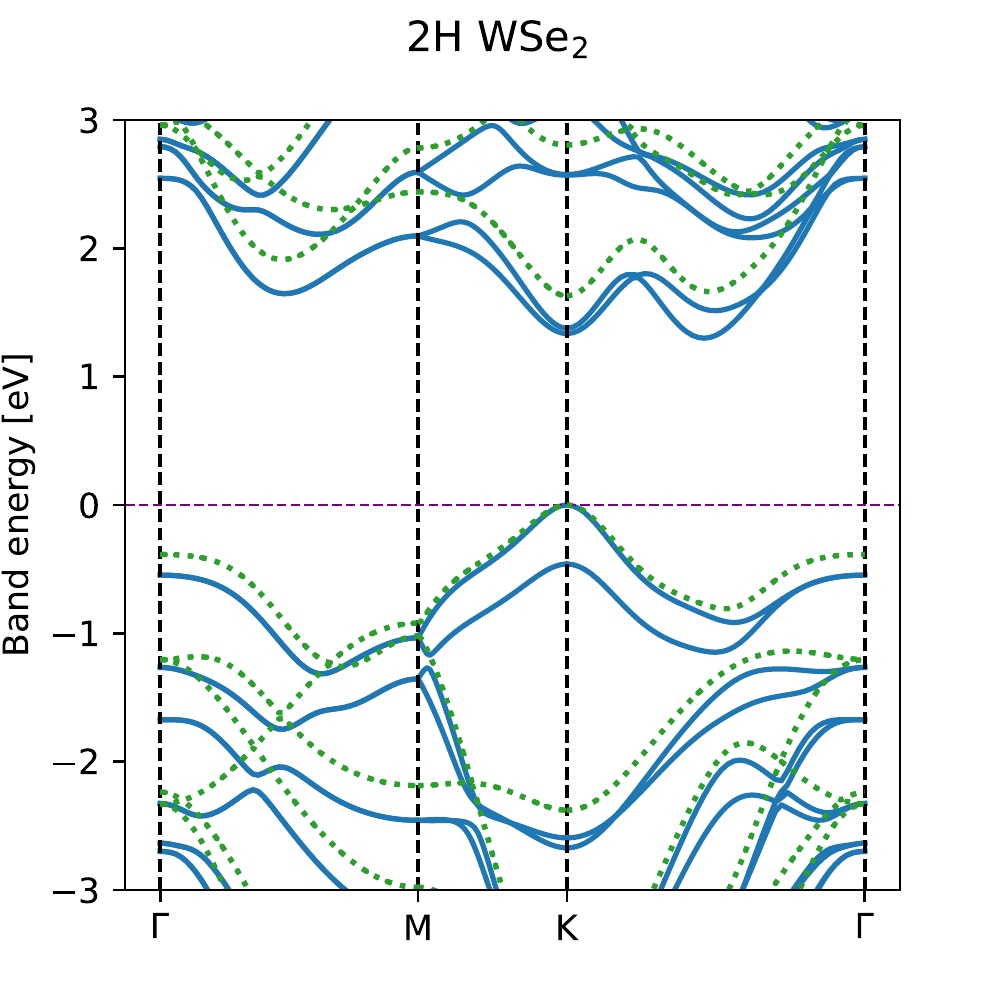}
\includegraphics[width=0.32\textwidth]{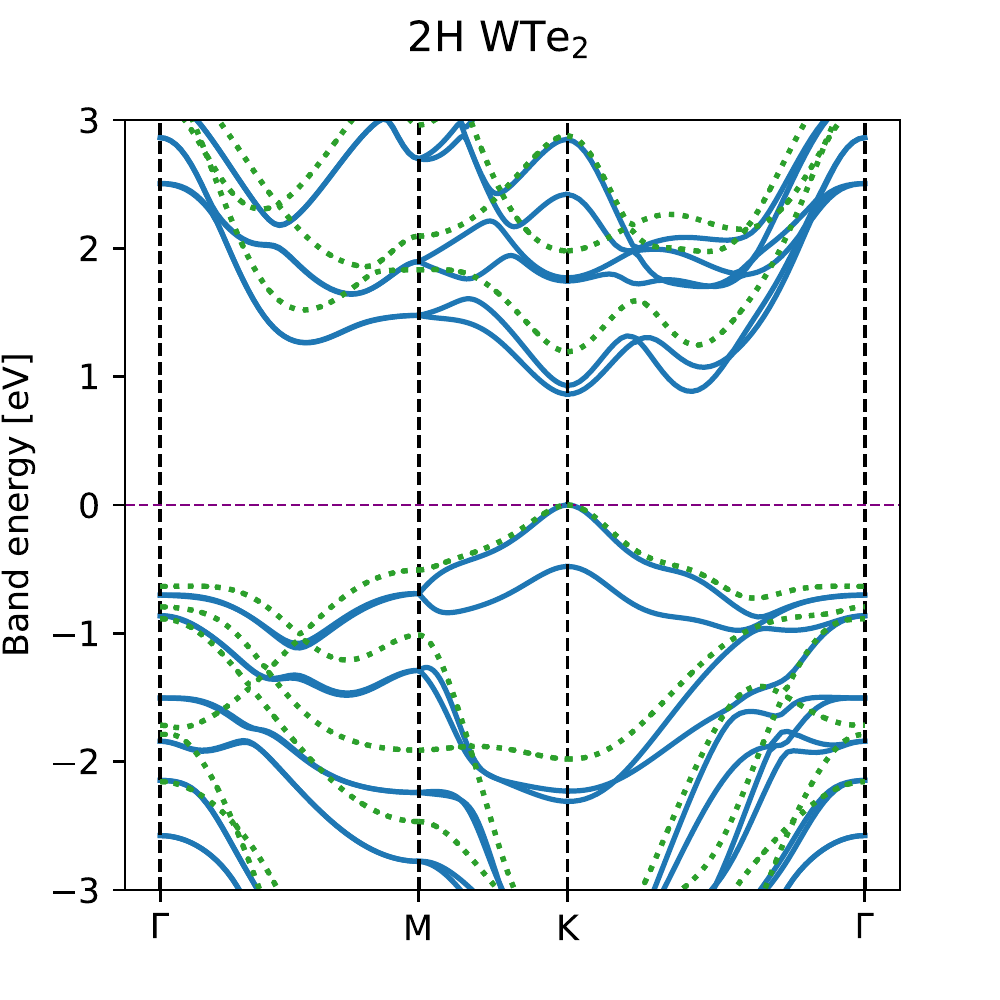}
\caption{\label{fig:mx2bsHex} Band structures of MX$_2$ monolayers in the 2H phase obtained from the \textsc{ReSpect} code at the scalar-relativistic (dashed lines, no SOC) and fully-relativistic (full lines, with SOC) levels of the theory. The horizontal dashed-black line separates the occupied and vacant states, and is placed on the top of the valence band for insulating systems. The unit cell-structure and lattice parameters were taken from Ref.~\onlinecite{miro2014atlas}.}
\end{figure*}

\begin{figure*}[tb]
\includegraphics[width=0.32\textwidth]{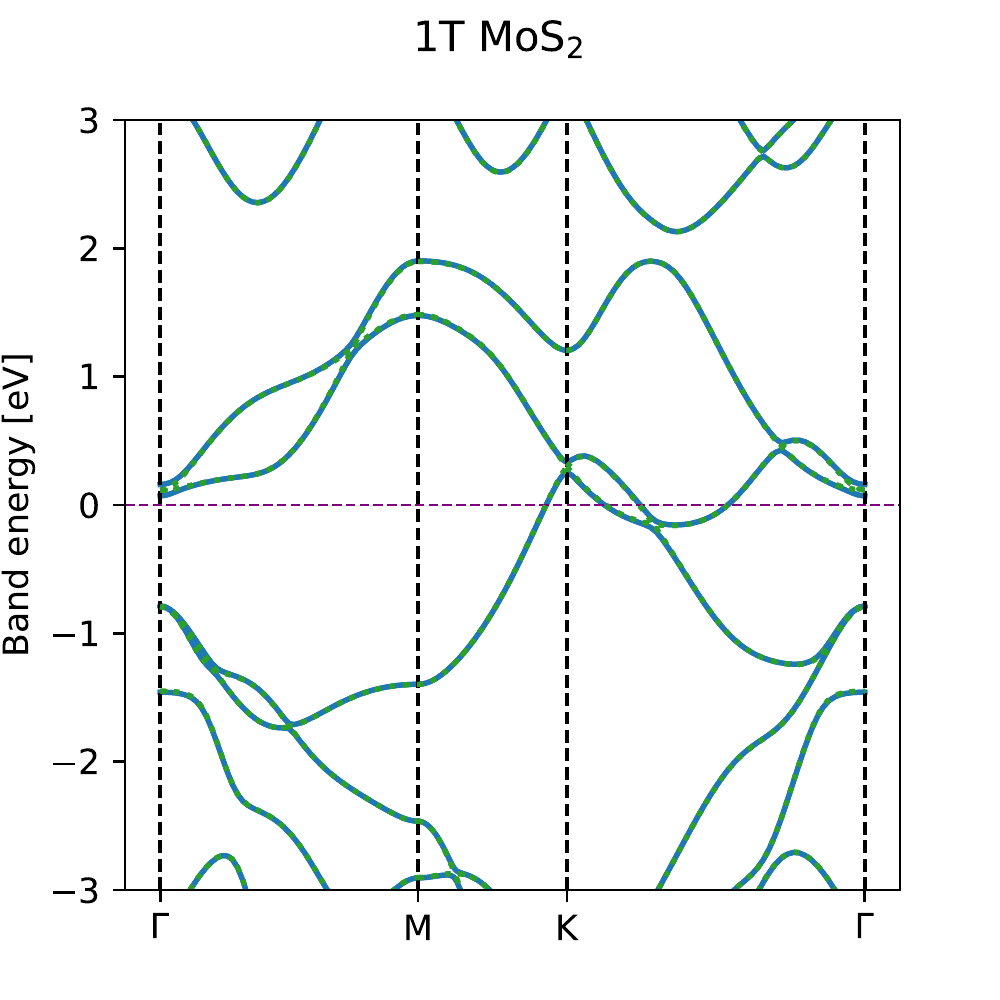}
\includegraphics[width=0.32\textwidth]{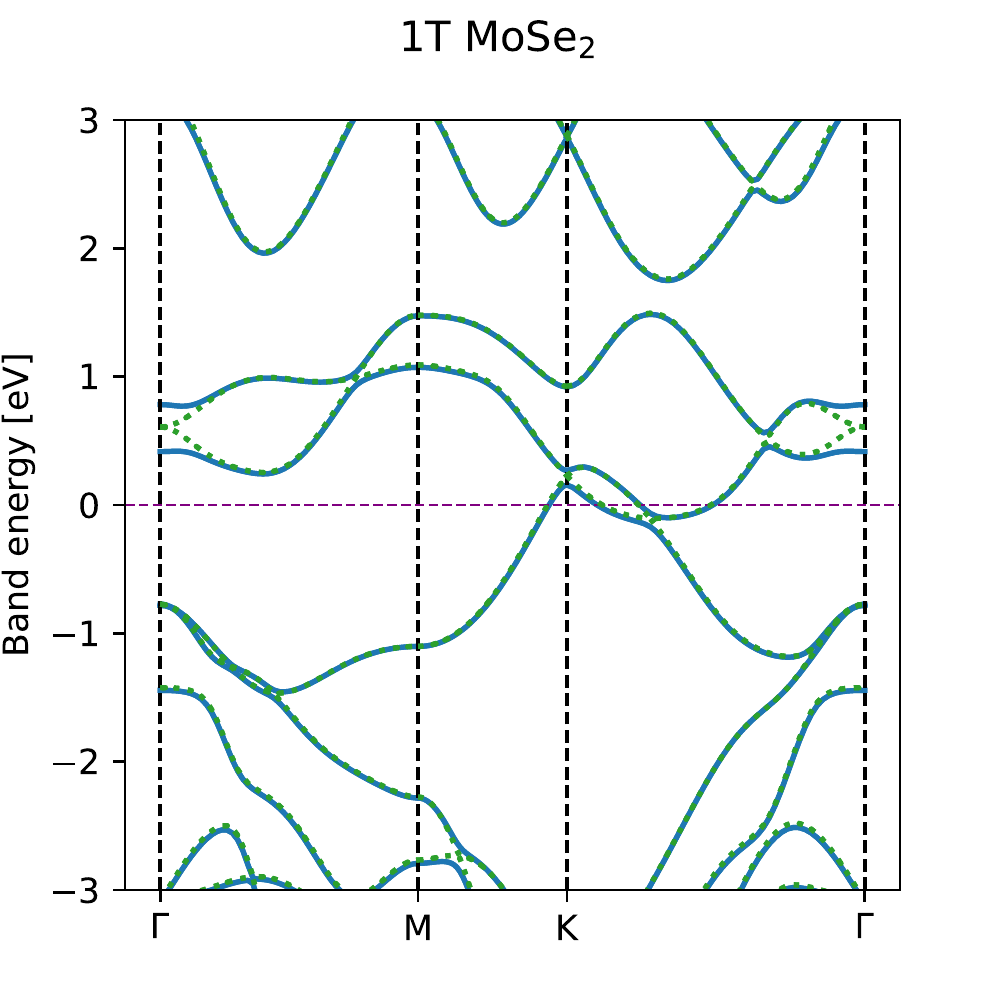}
\includegraphics[width=0.32\textwidth]{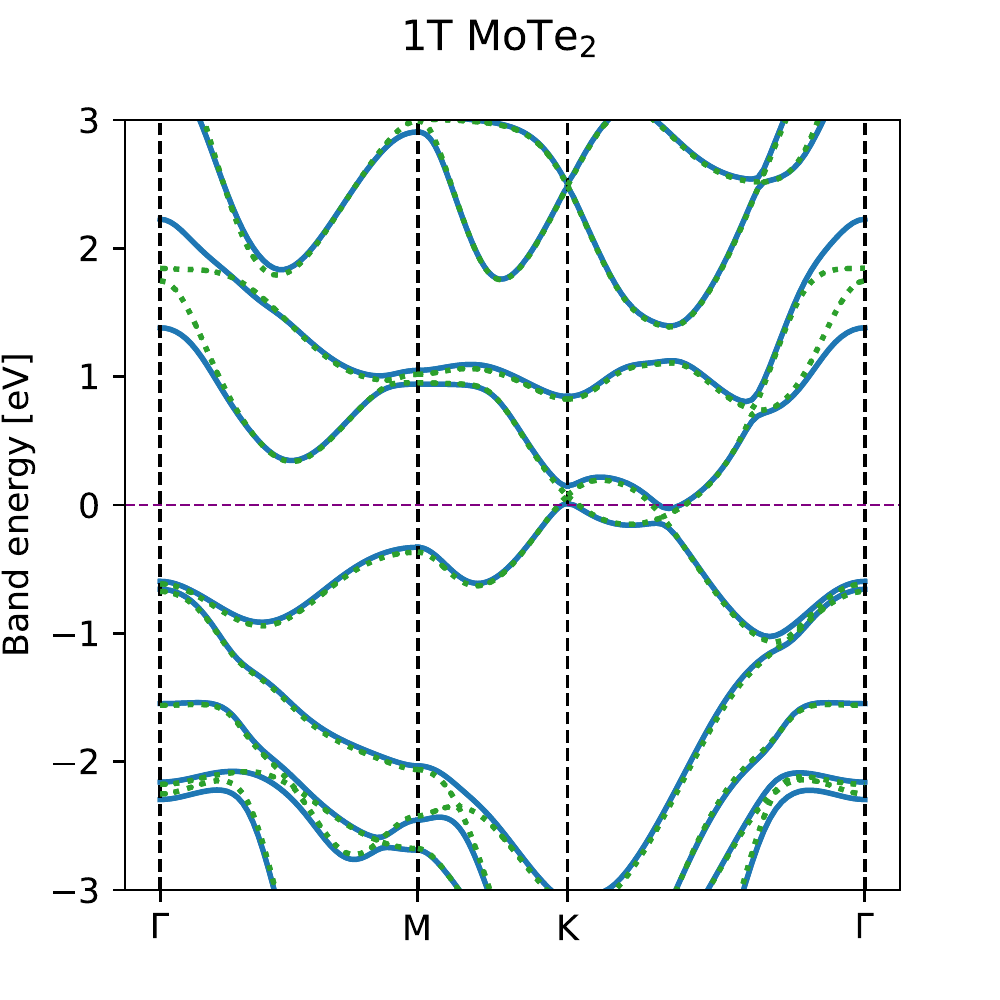}
\includegraphics[width=0.32\textwidth]{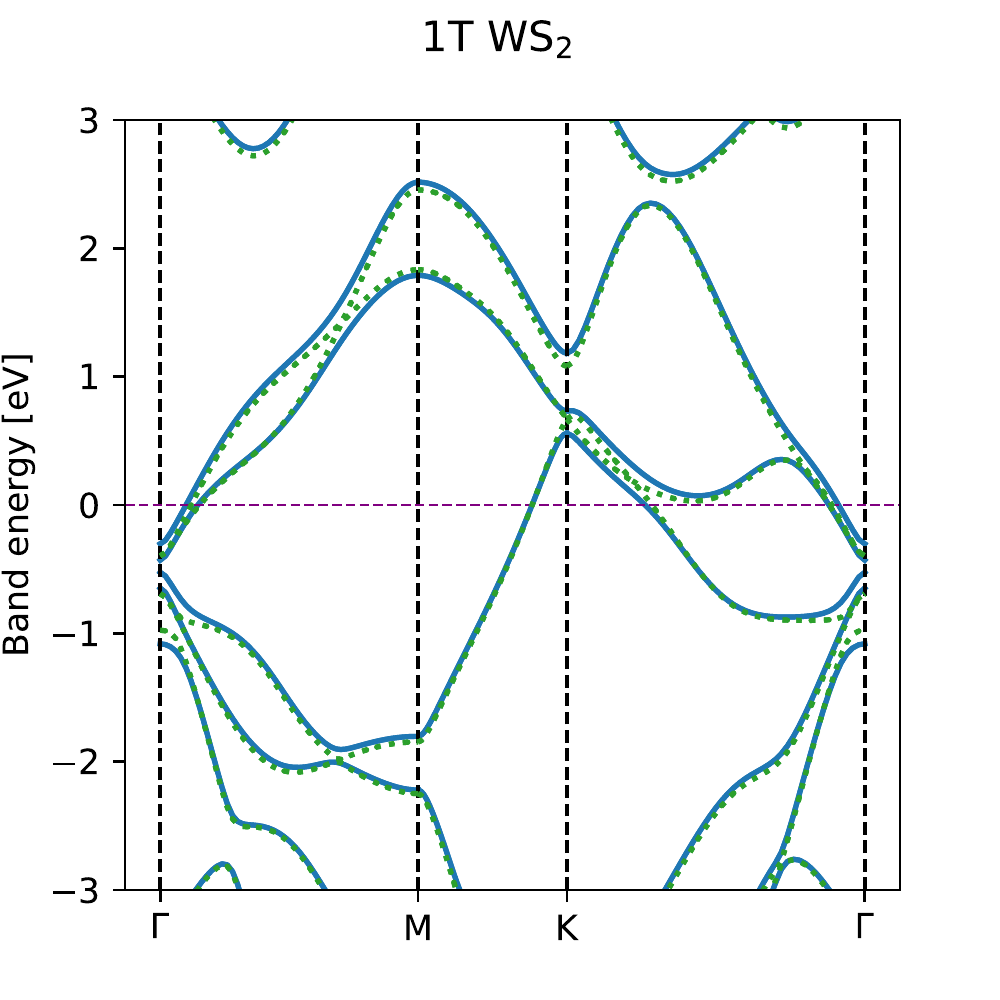}
\includegraphics[width=0.32\textwidth]{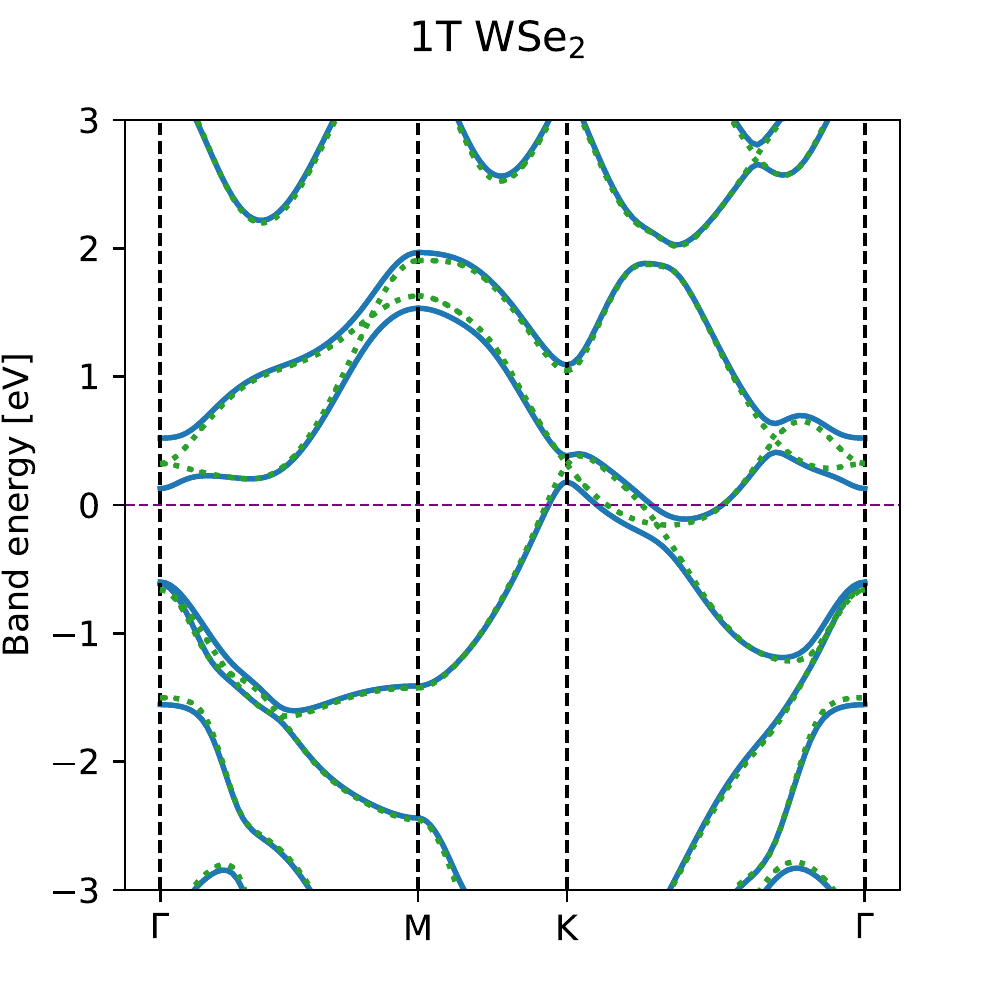}
\includegraphics[width=0.32\textwidth]{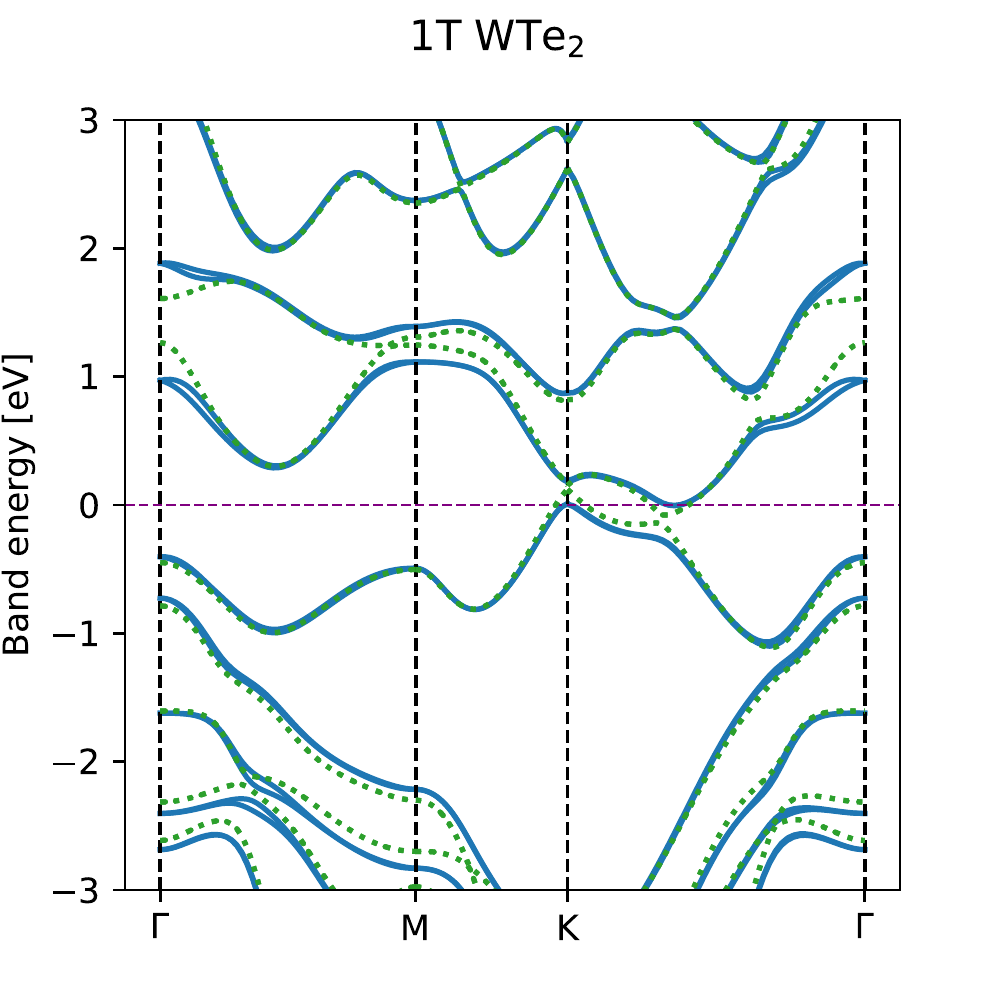}
\caption{\label{fig:mx2bsTet} Band structures of MX$_2$ monolayers in the 1T phase obtained from the \textsc{ReSpect} code at the scalar-relativistic (dashed lines, no SOC) and fully-relativistic (full lines, with SOC) levels of the theory. The horizontal dashed-black line marks the Fermi level. The unit-cell structure and lattice parameters were taken from Ref.~\onlinecite{miro2014atlas}.}
\end{figure*}

\begin{table}[tb]
\caption{\label{tab:gaps1T} Direct SOC-induced band gaps between the valence and conduction bands of MX$_2$ in the 1T phase at the $K$ symmetry point in the Brillouin zone.}
%\begin{ruledtabular}
\begin{tabular}{c @{\qquad} d}
\toprule
& \multicolumn{1}{c}{$K$-$K$ gap (eV)} \\
% \cline{2-3}
% & \multicolumn{1}{c}{sr} & \multicolumn{1}{c}{fr} \\
\colrule
% MoS$_2$  & 0.000  &  0.079  \\
% MoSe$_2$ & 0.000  &  0.114  \\
% MoTe$_2$ & 0.011  &  0.138  \\
% WS$_2$   & 0.000  &  0.175  \\
% WSe$_2$  & 0.000  &  0.205  \\
% WTe$_2$  & 0.032  &  0.173
MoS$_2$  & 0.079  \\
MoSe$_2$ & 0.114  \\
MoTe$_2$ & 0.138  \\
WS$_2$   & 0.175  \\
WSe$_2$  & 0.205  \\
WTe$_2$  & 0.173  \\
\botrule
\end{tabular}
%\end{ruledtabular}
\end{table}

\begin{figure*}[tb]
\includegraphics[width=0.32\textwidth]{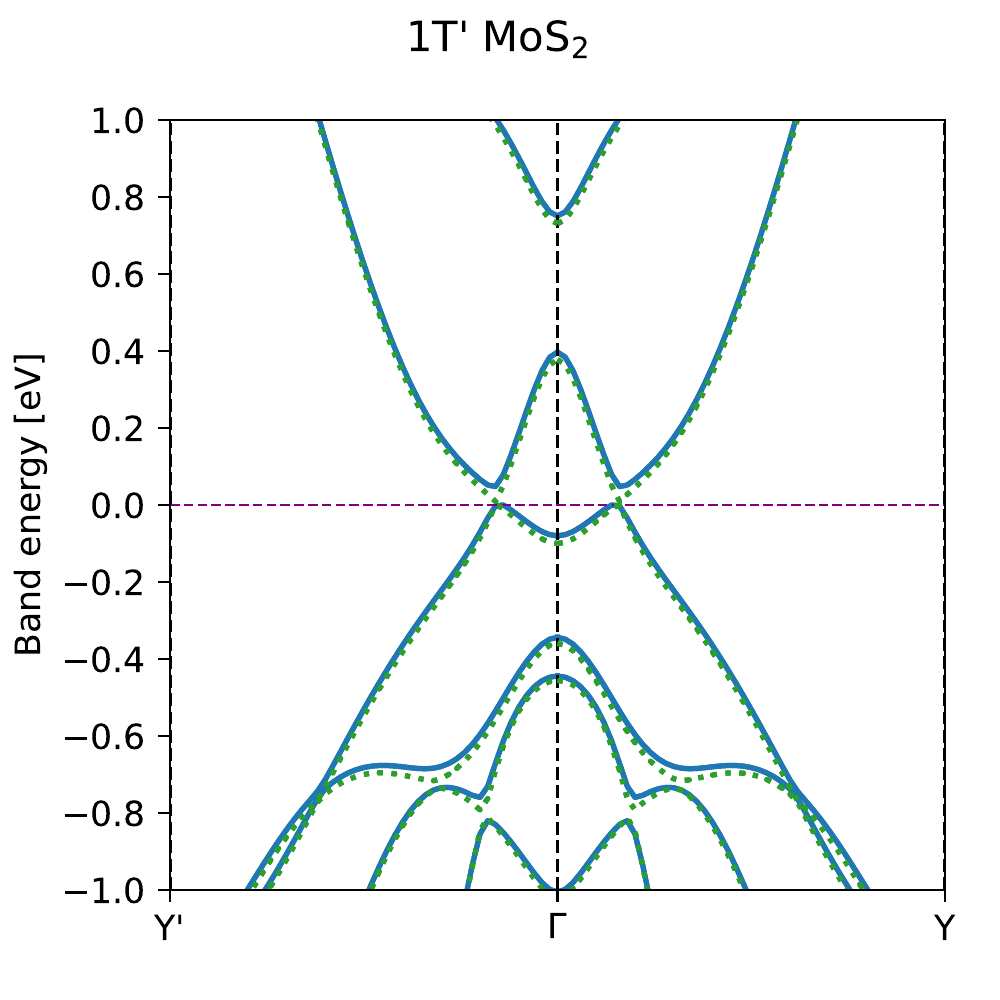}
\includegraphics[width=0.32\textwidth]{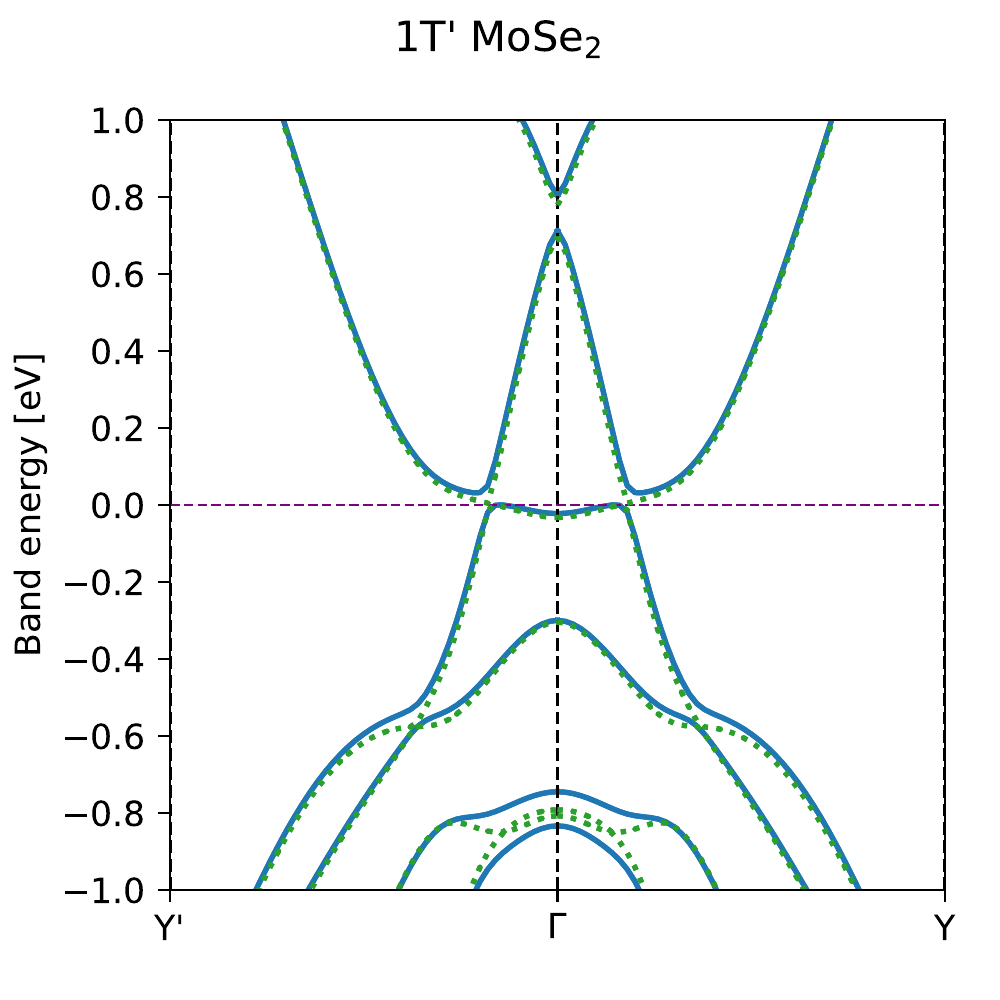}
\includegraphics[width=0.32\textwidth]{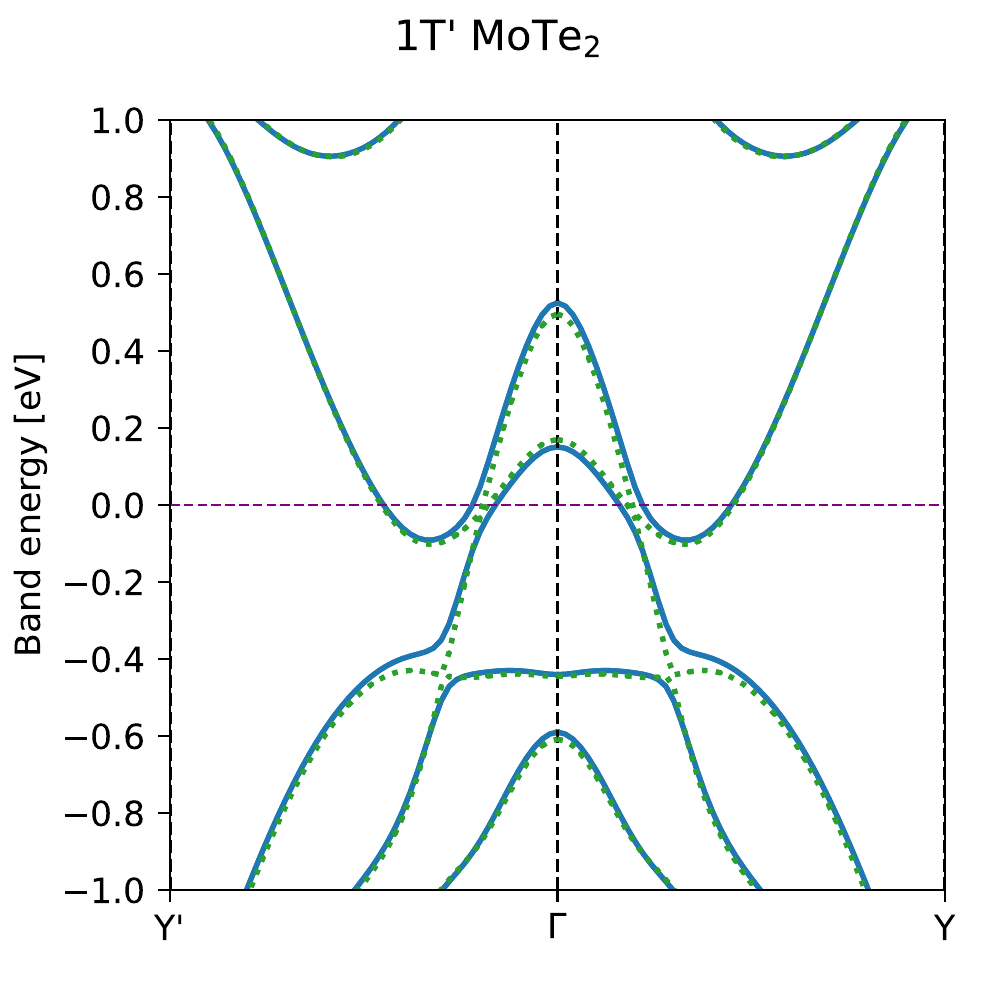}
\includegraphics[width=0.32\textwidth]{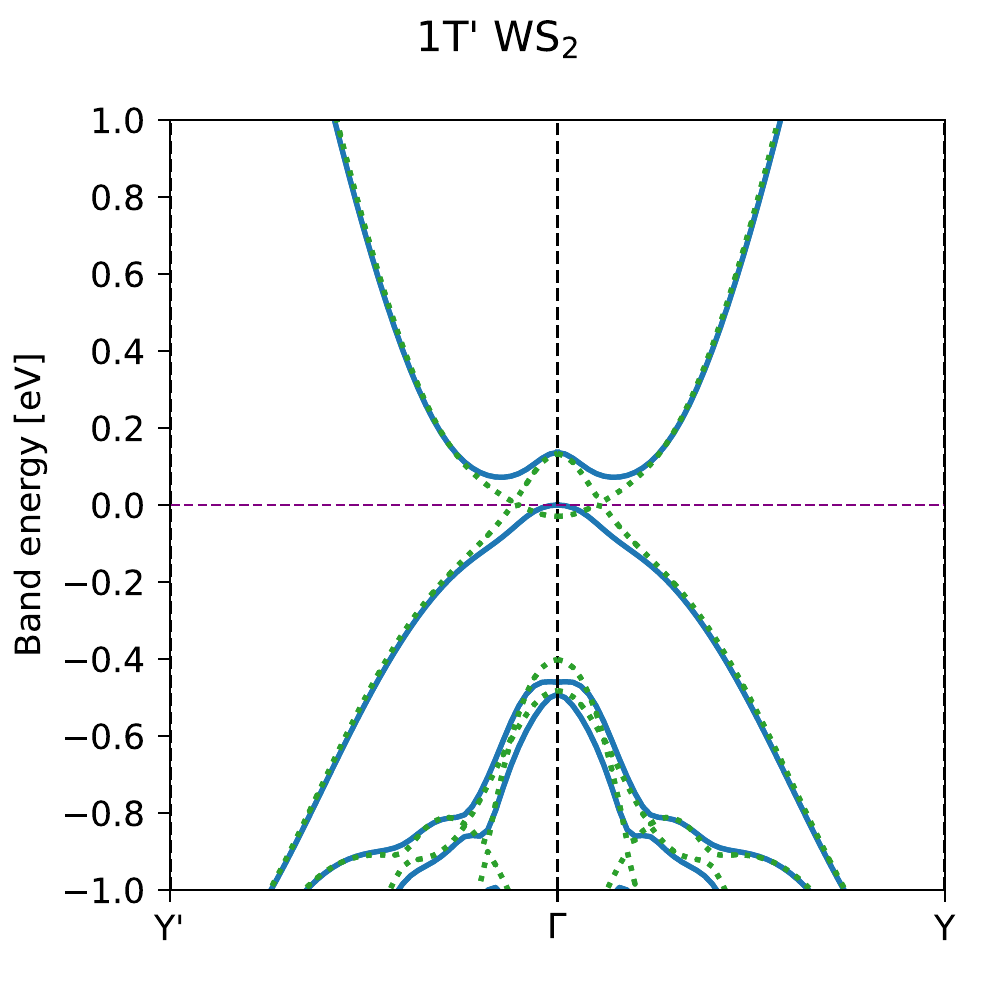}
\includegraphics[width=0.32\textwidth]{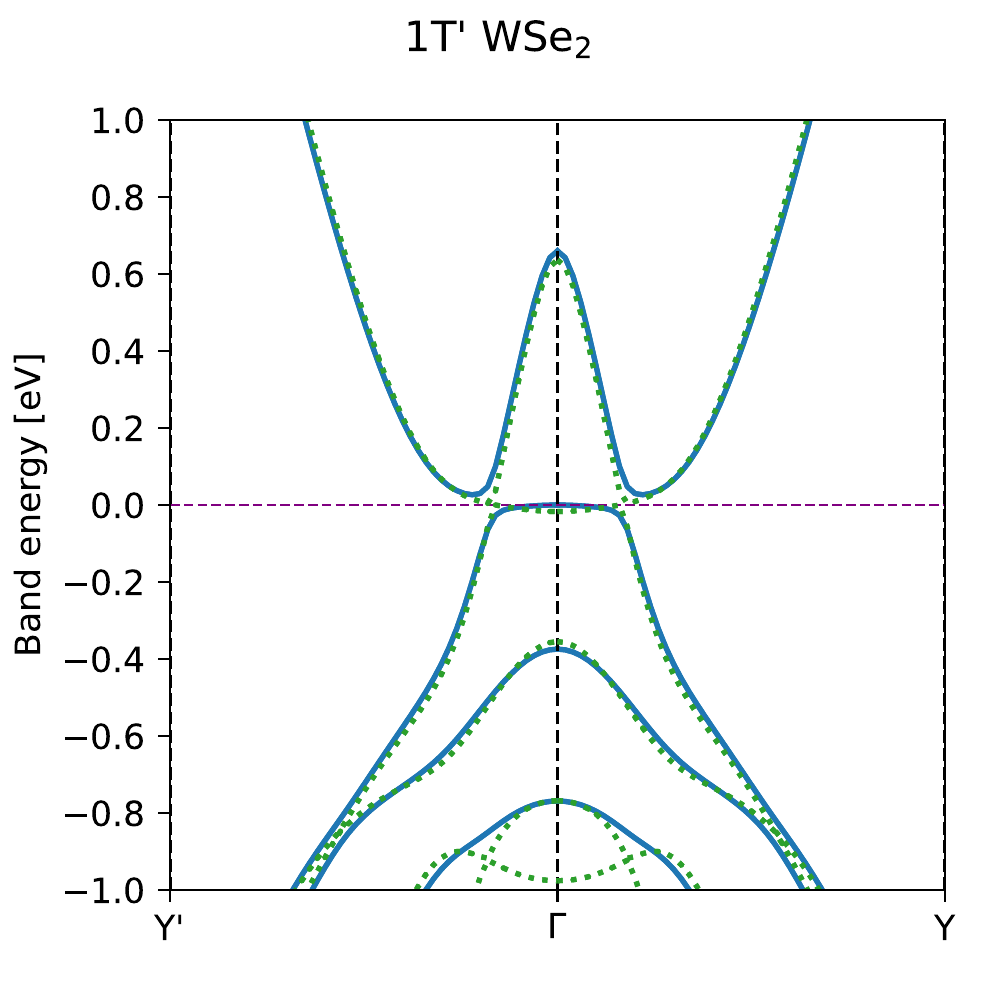}
\includegraphics[width=0.32\textwidth]{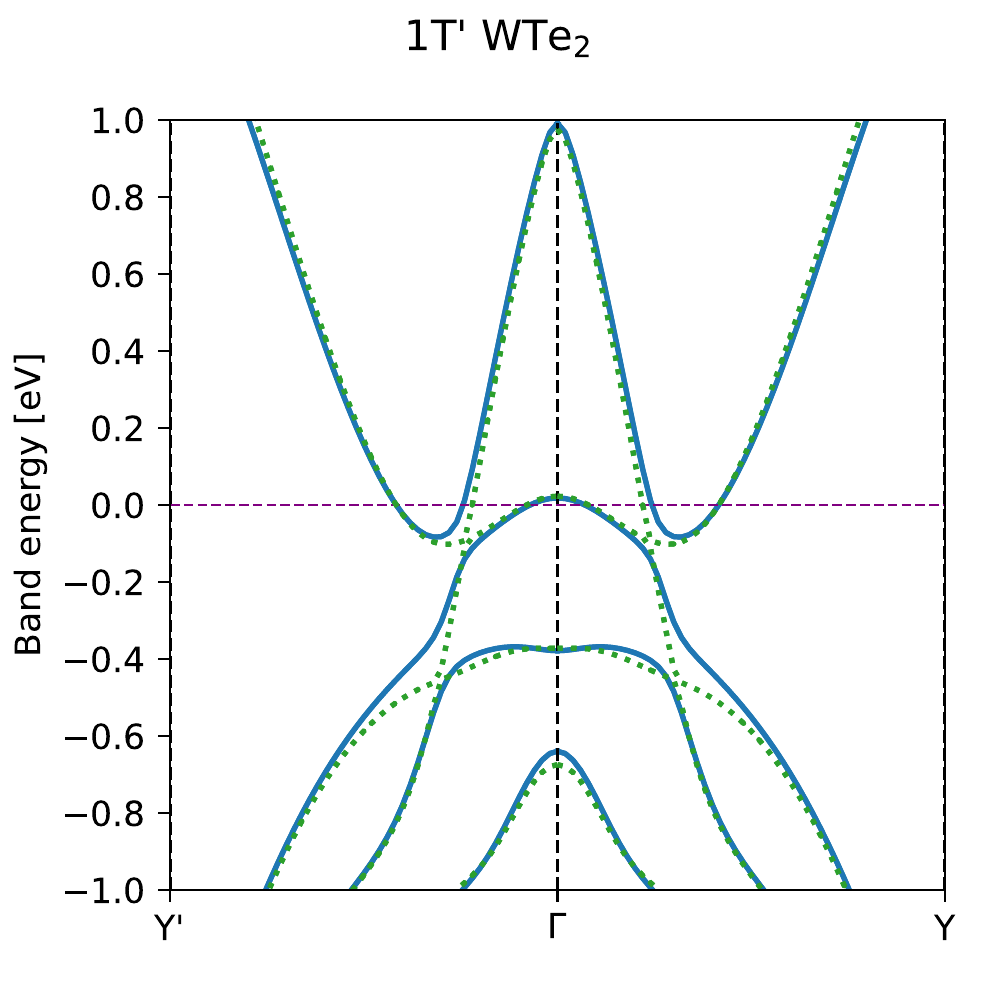}
\caption{\label{fig:mx2bsTetP} Band structures of MX$_2$ monolayers in the 1T' phase obtained from the \textsc{ReSpect} code at the scalar-relativistic (dashed lines, no SOC) and fully-relativistic (full lines, with SOC) levels of theory. The horizontal dashed-black line marks the Fermi level. The unit-cell structure and lattice parameters were taken from Ref.~\onlinecite{qian2014quantum}.}
\end{figure*}

\clearpage
\bibliography{references}% Produces the bibliography via BibTeX.